\documentclass[10pt, twocolumn]{IEEEtran}

\usepackage{graphicx}
\usepackage{color}

\usepackage{amsmath}
\usepackage{amssymb}
\usepackage{amsthm}
\usepackage{arydshln}    
\usepackage{bm}
\usepackage{cite}
\usepackage{mnsymbol}

\usepackage[caption=false,font=footnotesize]{subfig}
\usepackage[normalem]{ulem} 

\newtheorem{theorem}{Theorem}
\newtheorem{conjecture}[theorem]{Conjecture}

\newtheorem{lemma}[theorem]{Lemma}

\def\MID{\, | \,} 
\def\P{\mathbb{P}}
\def\Hhat{\widehat{H}}
\def\hhat{\widehat{h}}
\def\Belief{q}

\newcommand{\LocalISeq}[3]{{P_{e,#1_{#3}}^{{\rm I}_{#2}}}}    
\newcommand{\LocalIISeq}[3]{{P_{e,#1_{#3}}^{{\rm II}_{#2}}}}

\newcommand{\DecThres}[3]{{\lambda_{#1_{#3}}^{_{#2}}}}
\newcommand{\Frac}[2]{{{#1}}/{{#2}}}
\newcommand{\pFrac}[2]{{{(#1)}}/{{(#2)}}}

\definecolor{darkgreen}{RGB}{0,100,0}

\begin{document}

\title{Social Teaching: Being Informative vs. Being Right in Sequential Decision Making}

\author{Joong~Bum~Rhim and Vivek~K~Goyal%
\thanks{This material is based upon work supported by the National Science Foundation under Grant No.~1101147.}
\thanks{The authors are with the Research Laboratory of Electronics,
  Massachusetts Institute of Technology, Cambridge, MA 02139 USA.}
}

\maketitle

\begin{abstract}
We show that it can be suboptimal for Bayesian decision-making agents employing social learning to use correct prior probabilities as their initial beliefs.  We consider sequential Bayesian binary hypothesis testing where each individual agent makes a binary decision based on an initial belief, a private signal, and the decisions of all earlier-acting agents---with the actions of precedent agents causing updates of the initial belief.  Each agent acts to minimize Bayes risk, with all agents sharing the same Bayes costs for Type I (false alarm) and Type II (missed detection) errors.  The effect of the set of initial beliefs on the decision-making performance of the last agent is studied.  The last agent makes the best decision when the initial beliefs are inaccurate.  When the private signals are described by Gaussian likelihoods, the optimal initial beliefs are not haphazard but rather follow a systematic pattern:  the earlier-acting agents should act as if the prior probability is larger than it is in reality when the true prior probability is small, and vice versa.  We interpret this as being \emph{open minded} toward the unlikely hypothesis.  The early-acting agents face a trade-off between making a correct decision and being maximally informative to the later-acting agents.
\end{abstract}


\section{Introduction}
Consider decision-making agents
facing the task of choosing between two alternatives.
Each agent has a private signal, which is not visible to the other agents.
The agents sequentially make their individual decisions, which are visible to other agents.  An agent's action
contains some information about the right (or better) choice,
so subsequent agents can learn from the action
and reflect it in their own actions.
For example, when you want to choose between two alternatives when buying a new phone, the choices made by your colleagues can affect your judgment.

Being influenced by earlier-acting agents has been termed \emph{social learning}~\cite{EllisonF1993}.
It has generally been studied in settings where each agent has no motivation beyond making a correct choice for himself.
In this paper, we study the effect of an agent's action on subsequent agents and find that making correct decisions is generally not equivalent to providing information to other agents that maximally benefits them in their decision making tasks.  Accounting for the effect on other agents could be termed \emph{social teaching}.  Also, in any scenario with social learning, the earlier-acting agents can be seen as \emph{advisers} to the later-acting agents.  As will be detailed later, we find that a good adviser should be \emph{open minded} in the sense of being more receptive to the \emph{a priori} less likely alternative than she would have been if she were only interested in being right rather than also interested in being informative.

The framework of sequential decision making with social learning was independently introduced in~\cite{Banerjee1992} and~\cite{BikhchandaniHW1992}.  These works focused primarily on \emph{herding}, which is for all agents beyond some index to take the same action.  They showed that an incorrect herd would arise with positive probability when private signals are boundedly informative.\footnote{A signal $Y$ generated under a state $H$ is called \emph{boundedly informative} if there exists $\kappa > 0$ such that $\kappa < f_{Y \MID H}(y \MID h) < 1 / \kappa$ for all $y$ and $h$.}
For example, the private signals were assumed to be binary and to give true or false information, each with positive probability.
It can happen that a couple of the first agents receive false private signals and thus choose wrong actions. Then the effect of these actions on the beliefs of subsequent agents can be so great as to cause them to ignore their private signals and follow their precedent agents.  The private signals are bounded so that they cannot be strong enough to overcome the effect of the wrong actions.

Subsequently, \cite{SmithS2000} showed that learning is incomplete---beliefs are not eventually focused on the true state---if private signals are boundedly informative, but agents will asymptotically settle on the optimal action otherwise.
Recently, \cite{AcemogluDLO2011} extends the result to general network topologies where each agent can observe decisions made by its neighbors instead of all previous agents.

In another related line of work, \cite{Krishnamurthy2012} studies the effect of social learning in a quickest detection problem, in which agents keep updating their beliefs based on previous decisions and detect the time at which an underlying state changes.  It has a similar framework to \cite{HellmanCover1970}, which studied update of private information in a finite memory.

This paper differentiates itself from the literature in that it considers unbounded private signals and does not focus on herding behavior.
In addition,
we focus largely on the effect of prior probabilities in decision making.  We do not assume that an agent knows a correct prior probability for the decision at hand.  Even if he does, we do not assume he takes the shortsighted approach of using the prior probability only to optimize the correctness of his own decision.
Instead, we study the effect of the prior probability on the decisions of subsequent agents.
Sequential decision making is considered from a signal-processing perspective in~\cite{Varshney97} as well.  Its model is similar to ours except that there all agents know the true prior probability; this difference changes the problem substantially.

Our criterion for optimality is the Bayes risk of the final agent;  we assume a sequential decision making model in which only the decision made by the final agent matters.  Since sequential decision making is a hypothesis testing problem, agents adopt likelihood ratio tests to choose their actions~\cite{NeymanPearson1933}.  As they observe decisions or actions chosen by precedent agents, they compute or \emph{update} their beliefs to perform more precise likelihood ratio tests.  The update process depends on the initial prior belief and the history of decisions.  We derive a recursive belief update function.

Bayes-optimal agents need to know the prior probability in order to perform the likelihood ratio test.  Hence, intuitively, agents with wrong prior beliefs should degrade the decision making and yield higher Bayes risk.  In addition, they would misunderstand the public signals because they do not know others' beliefs.

Contrary to intuition, it turns out that wrong beliefs may improve the decision made by the final agent in sequential decision making.
Especially when the private signals are distorted by additive Gaussian noise, the optimal first agent is open minded:
He acts as if the prior probability is larger than it is in reality when the true prior probability is small, and vice versa.

Section~\ref{sec:Background} provides additional background and motivation from human decision makers.  Section~\ref{sec:ProblemDescription} describes our sequential decision making model.  In Section~\ref{sec:BeliefUpdate}, we investigate how agents interpret the decision history, update their beliefs, and make decisions according to their positions in the chain of agents. It is proven in Section~\ref{sec:OptimalInitialBelief} that the true prior probability is not the optimal prior belief for $N = 2$.  Examples for Gaussian likelihoods are presented in Section~\ref{sec:Example}.  Section~\ref{sec:Conclusion} concludes the paper.

\section{Background}    \label{sec:Background}

The mathematical model presented herein abstracts human decision makers so as to be broadly applicable.  We are motivated in part by a study about the correlation between a defendant's physical appearance and juror decisions~\cite{BrownHG2008}.  It states that jurors feel a defendant more intelligent when the defendant is wearing eyeglasses, which leads to fewer guilty verdicts.  It also says that wearing eyeglasses is especially effective for African-American defendants.  Several other studies have also revealed the importance of a defendant's physical appearance on a jury's decision making~\cite{Stewart1980,StephanTully1977,Efran1974}.  From a logical standpoint, a defendant's appearance should be irrelevant to judical decisions because eyeglasses have nothing to do with crimes, and we believe that jurors do their best to make fair and reasonable decisions based only on evidence.  Then why does this happen in reality?

We find the answer in Bayesian reasoning and the concept of prior probability.  Let us liken a jury trial to a hypothesis testing problem.  A defendant is metaphorically an object in one of two states: guilty or not guilty. Perceptions of evidence presented by a prosecutor or defense counsel are noisy observations about the defendant's true state.  Jurors are detectors that make a decision based upon the noisy observations.  However, one element of hypothesis testing is missing: the prior probability that the defendant commits a crime.

Reasonable human beings resemble Bayesian decision makers~\cite{Viscusi1985,SwetsTB1961,GlanzerHM2009,BraseCT1998}; they need to know the prior to compute the posterior probability.  The prior probability can be critical to the verdict when the evidence at trial is ambiguous.  The problem is that the jurors cannot know the defendant's true prior probability.  Hence, before reaching a verdict, the jurors judge the defendant's prior probability by how intelligent, how attractive, how friendly, and how threatening the defendant ``looks.''
They may be able to estimate the prior probability close to the true value but their estimates would not be the same as the defendant's true prior probability.

We are not defending or criticizing this phenomenon but just focusing on an interesting issue raised by it:  Human agents perform Bayesian hypothesis testing with inaccurate knowledge of prior probabilities.  While the prior probability is one of the basic elements of estimation,  
the effect of accuracy of the prior probability has not received a great deal of attention.
Initially building upon~\cite{VarshneyVarshney08}, we have previously studied the effect of categorization of problem instances as inducing quantization of prior probabilities~\cite{RhimVG2011b,RhimVG2012a,RhimVG2012c,RhimVG2012b}.
The present paper is more fundamental in that it addresses whether accurate prior probabilities are even the most favorable.  We have found that inaccurate perception of the prior probability may be beneficial in sequential decision making.

\section{Problem Description}    \label{sec:ProblemDescription}

\begin{figure}
    \centering
    \includegraphics[width=3.2in]{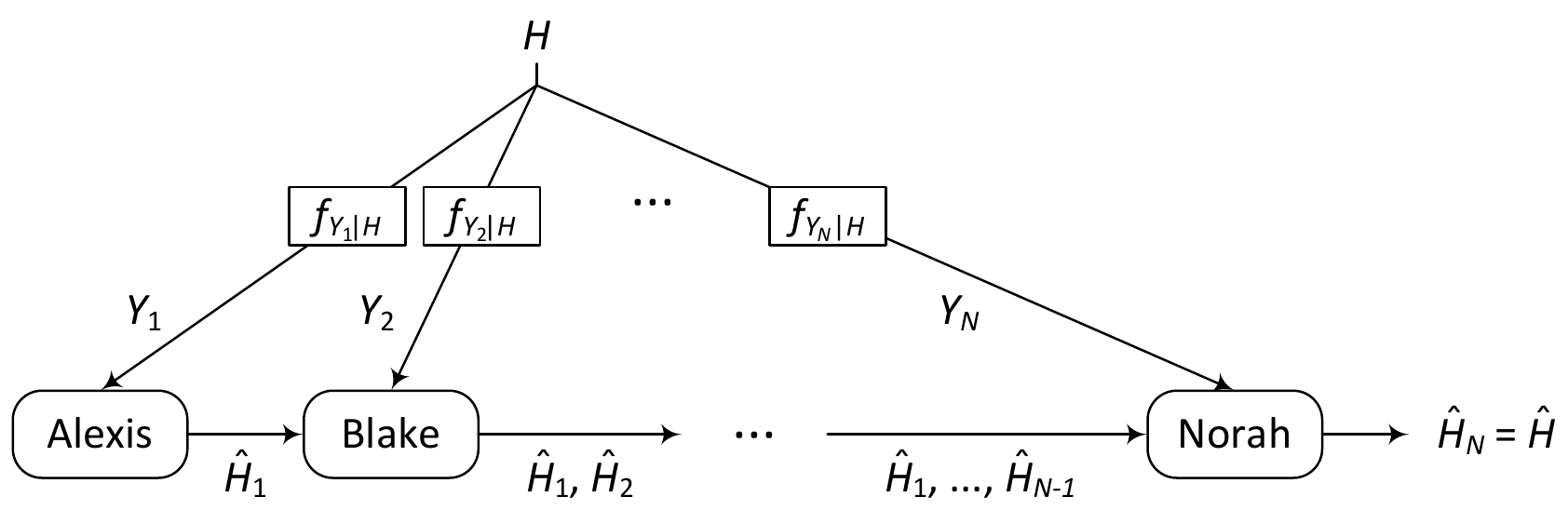}
    \caption{A sequential decision making model with $N$ agents (Alexis, Blake, \ldots, Norah).  The $n$th agent can observe $n - 1$ decisions made by the precedent agents.}
    \label{fig:SequentialDM}
\end{figure}

Consider the sequential decision making model depicted in Fig.~\ref{fig:SequentialDM}.  There is an object in a binary state $H \in \{0, 1\}$ with probability $\P(\{H = 0\}) = p_0$ and $\P(\{H = 1\}) = 1 - p_0$.
There are also $N$ agents that sequentially detect the state.  Agents do not know the true prior probability of the object.  Instead, the $n$th agent perceives it as $q_n$.  The $n$th agent observes decisions made by precedent agents, $\{\Hhat_1, \ldots, \Hhat_{n-1}\}$, as well as a signal about $H$, $Y_n$, generated from a likelihood function $f_{Y_n \MID H}$.\footnote{The signal in this model has a continuous value while the corrupted signal has a discrete value in the initially-proposed framework~\cite{Banerjee1992, BikhchandaniHW1992}.}  The private signals $\{Y_n\}$ are conditionally independent given $H$ and are identically distributed.  We assume that the likelihood ratio $f_{Y_n \MID H}(y_n \MID 1) / f_{Y_n \MID H}(y_n \MID 0)$ is an increasing function of $y_n$.

The $n$th agent can extract some information from the $n - 1$ precedent decisions.  The decisions are, however, biased by the false impressions that the precedent agents have of the object.  Even worse, the $n$th agent does not know what $q_1, \ldots, q_{n-1}$ are.  Thus, the $n$th agent assumes that they are all equal to $q_n$ and interprets the precedent decisions accordingly.
Using the history of decisions, the $n$th agent updates its prior belief before applying its likelihood ratio test.  We will define a recursive function that describes the belief update in Section~\ref{sec:BeliefUpdate}.\footnote{We use the term \emph{belief} to distinguish from the true prior probability and to capture that it is what agents believe as the prior probability.  Agents initially perceive the prior probability in some way, which we call the \emph{prior beliefs}.  After they observe precedent decisions, they modify the prior beliefs.  The beliefs are then called \emph{updated}.}

Our interest is in the last agent Norah and her decision $\Hhat_N$.  Upon observing her private signal $Y_N$ and the $N - 1$ precedent decisions $\Hhat_1, \ldots, \Hhat_{N-1}$, she determines her decision rule.  We evaluate the decision rule by a common criterion, Bayes risk, which measures the expected cost of her decision.  The relative importance of correct decisions and errors can be abstracted as a cost function $c(\Hhat, H)$, which defines penalties for false alarm or Type I error (choosing $\Hhat = 1$ when $H = 0$), correct rejection (choosing $\Hhat = 0$ when $H = 0$), hit (choosing $\Hhat = 1$ when $H = 1$), and missed detection or Type II error (choosing $\Hhat = 0$ when $H = 1$).  For simplicity, we assume the correct decisions have zero cost and use the shorthand notations $c_{10} = c(1, 0)$ and $c_{01} = c(0, 1)$ respectively for costs of false alarms and missed detections.  In addition, we consider agents have the same costs; they are a team in the sense of Radner~\cite{Radner1962}.  Then the Bayes risk is given by
\begin{equation}
    R_N = c_{10} p_0 p_{\Hhat_N \MID H}(1 \MID 0) + c_{01} (1 - p_0) p_{\Hhat_N \MID H}(0 \MID 1).
    \label{eq:BayesRisk1}
\end{equation}
The computation of \eqref{eq:BayesRisk1} depends on the previous decisions $\Hhat_1, \ldots, \Hhat_{N-1}$.  Therefore, the correct computation of the expected cost is
\begin{align}
    R_N = & \sum_{\hhat_1, \ldots, \hhat_{N-1}} \left( c_{10} p_0 p_{\Hhat_N, \Hhat_{N-1}, \ldots, \Hhat_1 \MID H}(1, \hhat_{N-1}, \ldots, \hhat_1 \MID 0) \right. \nonumber \\
    & \left. + c_{01} (1 - p_0) p_{\Hhat_N, \Hhat_{N-1}, \ldots, \Hhat_1 \MID H}(0, \hhat_{N-1}, \ldots, \hhat_1 \MID 1) \right).
    \label{eq:BayesRisk2}
\end{align}
We will discuss the optimal values of $q_n$ that minimize \eqref{eq:BayesRisk2}.

It is important to note that, in our model, each agent uses a decision rule optimized for her own belief; the agents do not adjust their decision rules for the sake of Norah.  In other words, for all $n = 1, \ldots, N$, the $n$th agent adopts the decision rule that minimizes her Bayes risk $R_n$, and her decision is shown to the other agents as a public signal.
In contrast, the agents could adjust their decision rules in an attempt to minimize the Bayes risk of a single collective decision, as studied for the combination of social learning and aggregation by voting in~\cite{RhimG:12arXiv-Ballot}.

We now introduce additional notation for the rest of the paper.  Random variables are in uppercase while their realizations are in lowercase.  We denote a probability density function (pdf) of a continuous random variable as $f$ and a probability mass function (pmf) of a discrete random variable as $p$.
A subscript number $n$ means ``of the $n$th agent.''  Superscript alphabet A (B) means ``upon observing Alexis's (Blake's) decision''; we sometimes use 0 or 1 instead of the Roman alphabet to specify a decision value.  For example, $\Belief_3^{_\text{AB}}$ denotes the updated belief of the third agent, Chuck, upon observing Alexis's and Blake's decisions $\Hhat_1$ and $\Hhat_2$, and $\Belief_3^{_{10}}$ denotes Chuck's updated belief upon observing $\Hhat_1 = 1$ and $\Hhat_2 = 0$. Subscript alphabet A (B) means ``that Alexis (Blake) thinks.''  For example, Blake thinks that the probability of Alexis choosing 0 when the true state is 0 is $p_{\Hhat_1 \MID H}(0 \MID 0)_{_\text{B}}$.  We need to clarify who thinks it because the agents are not aware of others' prior beliefs.  This will be explained in detail in Section~\ref{sec:Blake}.

\section{Prior Belief Update and Decision Making}    \label{sec:BeliefUpdate}

Our model assumes unbounded private signals.  Thus, unlike in~\cite{Banerjee1992,BikhchandaniHW1992}, it is always possible that a subsequent agent may not follow previous decisions; that is, incorrect herding does not occur.
We now discuss the utilization of a decision history as well as private signals for Bayesian hypothesis testing.  A position-wise decision-making strategy will be provided.  This can be interpreted as each agent updating his prior belief based on the decision history and then applying a likelihood ratio test with his private signal.

\subsection{Alexis, the First Agent}
Alexis performs normal binary hypothesis testing because she has no precedent decision.  She use the following likelihood ratio test with her prior belief $q_1$:
\begin{equation}
    \frac{f_{Y_1 \MID H} (y_1 \MID 1)}{f_{Y_1 \MID H} (y_1 \MID 0)} \overset{\Hhat_1(y_1) = 1}{\underset{\Hhat_1(y_1) = 0}{\gtreqless}}  \frac{c_{10} q_1}{c_{01} (1 - q_1)}.
    \label{eq:DMRule1}
\end{equation}
Since the likelihood ratio is increasing in $y_1$, the likelihood ratio test can be simplified to comparison with an appropriate decision threshold:
\begin{equation}
    y_1 \overset{\Hhat_1(y_1) = 1}{\underset{\Hhat_1(y_1) = 0}{\gtreqless}} \DecThres{}{}{}(q_1),
\end{equation}
where $\lambda(q)$ denotes the decision threshold that satisfies
\begin{equation}
    \frac{f_{Y \MID H} (\lambda \MID 1)}{f_{Y \MID H} (\lambda \MID 0)} = \frac{c_{10} q}{c_{01} (1 - q)}.
    \label{eq:Threshold}
\end{equation}

\subsection{Blake, the Second Agent}    \label{sec:Blake}
Blake observes Alexis's decision.  Considering $\Hhat_1$ as another corrupted signal of $H$ like $Y_2$, he modifies the likelihood function \eqref{eq:DMRule1} with his prior belief $q_2$ to
\begin{equation}
    \frac{f_{Y_2, \Hhat_1 \MID H} (y_2, \hhat_1 \MID 1)}{f_{Y_2, \Hhat_1 \MID H} (y_2, \hhat_1 \MID 0)} \overset{\Hhat_2(y_2) = 1}{\underset{\Hhat_2(y_2) = 0}{\gtreqless}}  \frac{c_{10} \Belief_2}{c_{01} (1 - \Belief_2)}.
    \label{eq:DMRule2}
\end{equation}
In the left-hand side of \eqref{eq:DMRule2},
\begin{displaymath}
f_{Y_2, \Hhat_1 \MID H} (y_2, \hhat_1 \MID h) = f_{Y_2 \MID H} (y_2 \MID h) \, p_{\Hhat_1 \MID H} (\hhat_1 \MID h)
\end{displaymath}
because the private signals $Y_1$ and $Y_2$ are conditionally independent given $H$.  We can rewrite \eqref{eq:DMRule2} as\footnote{The subscript ``B'' in the term ``$p_{\Hhat_1 \MID H}(\hhat_1 \MID h)_{_\text{B}}$'' indicates the value of $p_{\Hhat_1 \MID H}(\hhat_1 \MID h)$ that Blake (the second agent) thinks.  We specify this because Blake does not know Alexis's belief $q_1$.  Thus, he interprets her decision based on his belief $q_2$.  The value is different from the true value of $p_{\Hhat_1 \MID H}(\hhat_1 \MID h) = p_{\Hhat_1 \MID H}(\hhat_1 \MID h)_{_\text{A}}$.  Of course, it will be also different from what Chuck---the third agent---thinks, which is denoted by $p_{\Hhat_1 \MID H}(\hhat_1 \MID h)_{_\text{C}}$.  This will be explained in the next paragraph.}
\begin{equation}
    \frac{f_{Y_2 \MID H} (y_2 \MID 1)}{f_{Y_2 \MID H} (y_2 \MID 0)} \overset{\Hhat_2(y_2) = 1}{\underset{\Hhat_2(y_2) = 0}{\gtreqless}}  \frac{c_{10} \Belief_2}{c_{01} (1 - \Belief_2)} \frac{p_{\Hhat_1 \MID H}(\hhat_1 \MID 0)_{_\text{B}}}{p_{\Hhat_1 \MID H}(\hhat_1 \MID 1)_{_\text{B}}}.
    \label{eq:DMRule2-1}
\end{equation}

The likelihood ratio test \eqref{eq:DMRule2-1} can be interpreted with Blake updating his prior belief upon observing Alexis's decision $\Hhat_1$.  Combined with $\Belief_2$, $p_{\Hhat_1 \MID H}(\hhat_1 \MID h)_2$ updates his prior belief from $\Belief_2$ to $\Belief^{_\text{A}}_2$:
\begin{equation}
    \frac{\Belief^{_\text{A}}_2}{1 - \Belief^{_\text{A}}_2} = \frac{\Belief_2}{1 - \Belief_2} \frac{p_{\Hhat_1 \MID H}(\hhat_1 \MID 0)_{_\text{B}}}{p_{\Hhat_1 \MID H}(\hhat_1 \MID 1)_{_\text{B}}}.
    \label{eq:BeliefUpdate2-1}
\end{equation}
The updated belief is
\begin{align}
    \Belief^{_\text{A}}_2 & = \frac{\Belief_2 p_{\Hhat_1 \MID H}(\hhat_1 \MID 0)_{_\text{B}}}{\Belief_2 p_{\Hhat_1 \MID H}(\hhat_1 \MID 0)_{_\text{B}} + (1 - \Belief_2) p_{\Hhat_1 \MID H}(\hhat_1 \MID 1)_{_\text{B}}} \nonumber \\
    & = \frac{p_{\Hhat_1, H}(\hhat_1, 0)_{_\text{B}}}{p_{\Hhat_1, H}(\hhat_1, 0)_{_\text{B}} + p_{\Hhat_1, H}(\hhat_1, 1)_{_\text{B}}} \nonumber \\
    & = p_{H \MID \Hhat_1}(0 \MID \hhat_1)_{_\text{B}}.
    \label{eq:BeliefUpdate2}
\end{align}

We have to make clear that Blake does not correctly compute $p_{\Hhat_1 \MID H}(\hhat_1 \MID h)_{_\text{B}}$
because he does not know $\Belief_1$.
The true probability is given by
\begin{align}
    p_{\Hhat_1 \MID H}(0 \MID h) & = p_{\Hhat_1 \MID H}(0 \MID h)_{_\text{A}} = \P(\{Y_1 \leq \lambda(\Belief_1) \MID H = h\}) \nonumber \\
    & = \int_{-\infty}^{\lambda(\Belief_1)} f_{Y \MID H}(y \MID h) \, dy,
\end{align}
but Blake evaluates Alexis's decision $\Hhat_1$ as if it were made based on $\Belief_2$ not $\Belief_1$:
\begin{align}
    p_{\Hhat_1 \MID H}(0 \MID h)_{_\text{B}} & = \P(\{Y_1 \leq \lambda(\Belief_2) \MID H = h\}) \nonumber \\
    & = \int_{-\infty}^{\lambda(\Belief_2)} f_{Y \MID H}(y \MID h) \, dy.
    \label{eq:DecisionProbability2}
\end{align}

An interesting observation is that Alexis's biased belief $\Belief_1$ does not affect Blake's belief update.  There is no trace of $\Belief_1$ in \eqref{eq:BeliefUpdate2} and \eqref{eq:DecisionProbability2}.  Suppose that Alexis knows true prior probability $p_0$ and uses the decision threshold $\lambda(p_0)$.  Still Blake, who does not know what belief Alexis has, thinks that the conditional probability of Alexis declaring $\Hhat_1 = 0$ is given by \eqref{eq:DecisionProbability2} and updates his belief as in \eqref{eq:BeliefUpdate2}.  It is clear in \eqref{eq:BeliefUpdate2} that the updated belief depends only on Blake's initial belief and Alexis's decision.

However, Alexis's prior belief still affects Blake's performance in some way, which is related to the probability of error.  Alexis's biased belief changes the probability of her decision.  The changed probability is embedded in the probability of Blake's decision:
\begin{align*}
    p_{\Hhat_2 \MID H}(\hhat_2 \MID 0) & = \sum_{\hhat_1} p_{\Hhat_2, \Hhat_1 \MID H}(\hhat_2, \hhat_1 \MID 0) \nonumber \\
    & = p_{\Hhat_2 \MID \Hhat_1, H}(\hhat_2 \MID 0, 0)_{_\text{B}} \times p_{\Hhat_1 \MID H}(0 \MID 0)_{_\text{A}} \nonumber \\
    & \quad + p_{\Hhat_2 \MID \Hhat_1, H}(\hhat_2 \MID 1, 0)_{_\text{B}} \times p_{\Hhat_1 \MID H}(1 \MID 0)_{_\text{A}}, \nonumber \\
    p_{\Hhat_2 \MID H}(\hhat_2 \MID 1) & = \sum_{\hhat_1} p_{\Hhat_2, \Hhat_1 \MID H}(\hhat_2, \hhat_1 \MID 1) \nonumber \\
    & = p_{\Hhat_2 \MID \Hhat_1, H}(\hhat_2 \MID 0, 1)_{_\text{B}} \times p_{\Hhat_1 \MID H}(0 \MID 1)_{_\text{A}} \nonumber \\
    & \quad + p_{\Hhat_2 \MID \Hhat_1, H}(\hhat_2 \MID 1, 1)_{_\text{B}} \times p_{\Hhat_1 \MID H}(1 \MID 1)_{_\text{A}}.
\end{align*}
Thus, Alexis's biased belief changes the probability of Blake's decision as well as that of her decision.

\subsection{Chuck, the Third Agent}    \label{sec:Chuck}

Chuck's detection process is the same as Blake's.  He observes both Alexis's and Blake's decisions and also updates his prior belief $\Belief_3$ like in \eqref{eq:BeliefUpdate2-1}:
\begin{align*}
    \frac{\Belief^{_\text{AB}}_3}{1 - \Belief^{_\text{AB}}_3} & = \frac{\Belief_3}{1 - \Belief_3} \frac{p_{\Hhat_2, \Hhat_1 \MID H}(\hhat_2, \hhat_1 \MID 0)_{_\text{C}}}{p_{\Hhat_2, \Hhat_1 \MID H}(\hhat_2, \hhat_1 \MID 1)_{_\text{C}}} \nonumber \\
    & = \left( \frac{\Belief_3}{1 - \Belief_3} \frac{p_{\Hhat_1 \MID H}(\hhat_1 \MID 0)_{_\text{C}}}{p_{\Hhat_1 \MID H}(\hhat_1 \MID 1)_{_\text{C}}} \right) \frac{p_{\Hhat_2 \MID \Hhat_1, H}(\hhat_2 \MID \hhat_1, 0)_{_\text{C}}}{p_{\Hhat_2 \MID \Hhat_1, H}(\hhat_2 \MID \hhat_1, 1)_{_\text{C}}}.
\end{align*}
Be careful that $\Hhat_1$ and $\Hhat_2$ are not conditionally independent given $H$ because Blake's decision $\Hhat_2$ depends on Alexis's decision $\Hhat_1$:
\[p_{\Hhat_2 | \Hhat_1, H}(\hhat_2 \MID \hhat_1, 0) \neq p_{\Hhat_2 | H}(\hhat_2 \MID 0).\]

Chuck's update process can be split into in two steps.  The first step is to infer Blake's updated belief based on Alexis's decision:
\begin{align}
    \frac{\Belief^{_\text{A}}_3}{1 - \Belief^{_\text{A}}_3} = \frac{\Belief_3}{1 - \Belief_3} \ \frac{p_{\Hhat_1 \MID H}(\hhat_1 \MID 0)_{_\text{C}}}{p_{\Hhat_1 \MID H}(\hhat_1 \MID 1)_{_\text{C}}}.
    \label{eq:BeliefUpdate3-1}
\end{align}
The second step is to update his own belief from $\Belief_3^{_\text{A}}$ based on Blake's decision:
\begin{align}
    \frac{\Belief^{_\text{AB}}_3}{1 - \Belief^{_\text{AB}}_3} = \frac{\Belief^{_\text{A}}_3}{1 - \Belief^{_\text{A}}_3} \ \frac{p_{\Hhat_2 \MID \Hhat_1, H}(\hhat_2 \MID \hhat_1, 0)_{_\text{C}}}{p_{\Hhat_2 \MID \Hhat_1, H}(\hhat_2 \MID \hhat_1, 1)_{_\text{C}}}.
    \label{eq:BeliefUpdate3-2}
\end{align}
Please note that Chuck does not know Alexis's and Blake's prior beliefs, $\Belief_1$ and $\Belief_2$, like Blake did not know Alexis's.  Thus Chuck infers everything based on his own belief $\Belief_3$, which is indicated by the subscript ``C'' in \eqref{eq:BeliefUpdate3-1} and \eqref{eq:BeliefUpdate3-2}.

Details of computations of \eqref{eq:BeliefUpdate3-1} and \eqref{eq:BeliefUpdate3-2} are as follows:
\begin{subequations}
    \label{eq:DecisionProbability3-1}
\begin{align}
    p_{\Hhat_1 \MID H}(0 \MID h)_{_\text{C}} & 
     = \int_{-\infty}^{\lambda(\Belief_3)} f_{Y_1 \MID H}(y \MID h) \, dy, \\
    p_{\Hhat_1 \MID H}(1 \MID h)_{_\text{C}} & 
     = \int_{\lambda(\Belief_3)}^{\infty} f_{Y_1 \MID H}(y \MID h) \, dy.
\end{align}
\end{subequations}
Substituting \eqref{eq:DecisionProbability3-1} in \eqref{eq:BeliefUpdate3-1}, Chuck can compute $\Belief^{_\text{A}}_3$ for $\Hhat_1 = 0$ and $\Hhat_1 = 1$ respectively:
\begin{subequations}
    \label{eq:BeliefUpdate3-11}
\begin{align}
    \Belief^{_0}_3 & = \frac{\Belief_3}{\Belief_3 + (1 - \Belief_3) \frac{\int_{-\infty}^{\lambda(\Belief_3)} f_{Y_1 \MID H}(y \MID 1) \, dy}{\int_{-\infty}^{\lambda(\Belief_3)} f_{Y_1 \MID H}(y \MID 0) \, dy}}, \\
    \Belief^{_1}_3 & = \frac{\Belief_3}{\Belief_3 + (1 - \Belief_3) \frac{\int_{\lambda(\Belief_3)}^{\infty} f_{Y_1 \MID H}(y \MID 1) \, dy}{\int_{\lambda(\Belief_3)}^{\infty} f_{Y_1 \MID H}(y \MID 0) \, dy}}.
\end{align}
\end{subequations}
Then,
\begin{subequations}
    \label{eq:DecisionProbability3-2}
\begin{align}
    p_{\Hhat_2 \MID \Hhat_1, H}(0 \MID \hhat_1, h)_3 & = \P(\{Y_2 \leq \lambda(\Belief^{_\text{A}}_3) \MID H = h\}) \nonumber \\
    & = \int_{-\infty}^{\lambda(\Belief^{_\text{A}}_3)} f_{Y_2 \MID H}(y \MID h) \, dy, \\
    p_{\Hhat_2 \MID \Hhat_1, H}(1 \MID \hhat_1, h)_3 & = \P(\{Y_2 > \lambda(\Belief^{_\text{A}}_3) \MID H = h\}) \nonumber \\
    & = \int_{\lambda(\Belief^{_\text{A}}_3)}^{\infty} f_{Y_2 \MID H}(y \MID h) \, dy.
\end{align}
\end{subequations}
Even though the value of $\hhat_1$ may not seem to be used in \eqref{eq:DecisionProbability3-2}, it is inherent in $\Belief^{_\text{A}}_3$ and affects the computation results.
Chuck's updated belief $\Belief^{_\text{AB}}_3$ is obtained by substituting \eqref{eq:BeliefUpdate3-11} and \eqref{eq:DecisionProbability3-2} in \eqref{eq:BeliefUpdate3-2}.

\subsection{Norah, the $N$th Agent}
Norah, the $N$th agent, observes $Y_N$ and $\Hhat_1, \ldots, \Hhat_{N-1}$.  Paralleling the arguments in the preceding sections, her prior belief update is a function of $\Belief_N$ as well as $\Hhat_1, \ldots, \Hhat_{N-1}$ but not of $\Belief_1, \ldots, \Belief_{N-1}$.  Thus, we can define a general prior belief update function: \[\Belief^{_\text{AB$\cdots$M}}_N = U_N(\Belief_N, \hhat_1, \hhat_2, \ldots, \hhat_{N-1}).\]

The function $U_n$ has a recurrence relation:
\begin{itemize}
\item For $n = 1$, $U_1(q) = q$.
\item For $n > 1$,
\begin{align*}
     U_n(q, \hhat_1, \ldots, \hhat_{n-2}, 0) 
     & = \frac{\tilde{q}}{\tilde{q} + (1 - \tilde{q}) \frac{\int_{-\infty}^{\lambda(\tilde{q})} f_{Y_{n-1} \MID H}(y \MID 1) \, dy}{\int_{-\infty}^{\lambda(\tilde{q})} f_{Y_{n-1} \MID H}(y \MID 0) \, dy}}, \nonumber \\
    U_n(q, \hhat_1, \ldots, \hhat_{n-2}, 1) 
     & = \frac{\tilde{q}}{\tilde{q} + (1 - \tilde{q}) \frac{\int_{\lambda(\tilde{q})}^{\infty} f_{Y_{n-1} \MID H}(y \MID 1) \, dy}{\int_{\lambda(\tilde{q})}^{\infty} f_{Y_{n-1} \MID H}(y \MID 0) \, dy}},
\end{align*}
where $\tilde{q} = U_{n-1}(q, \hhat_1, \ldots, \hhat_{n-2})$.
\end{itemize}

\begin{figure}
    \centering
    \includegraphics[width=3.2in]{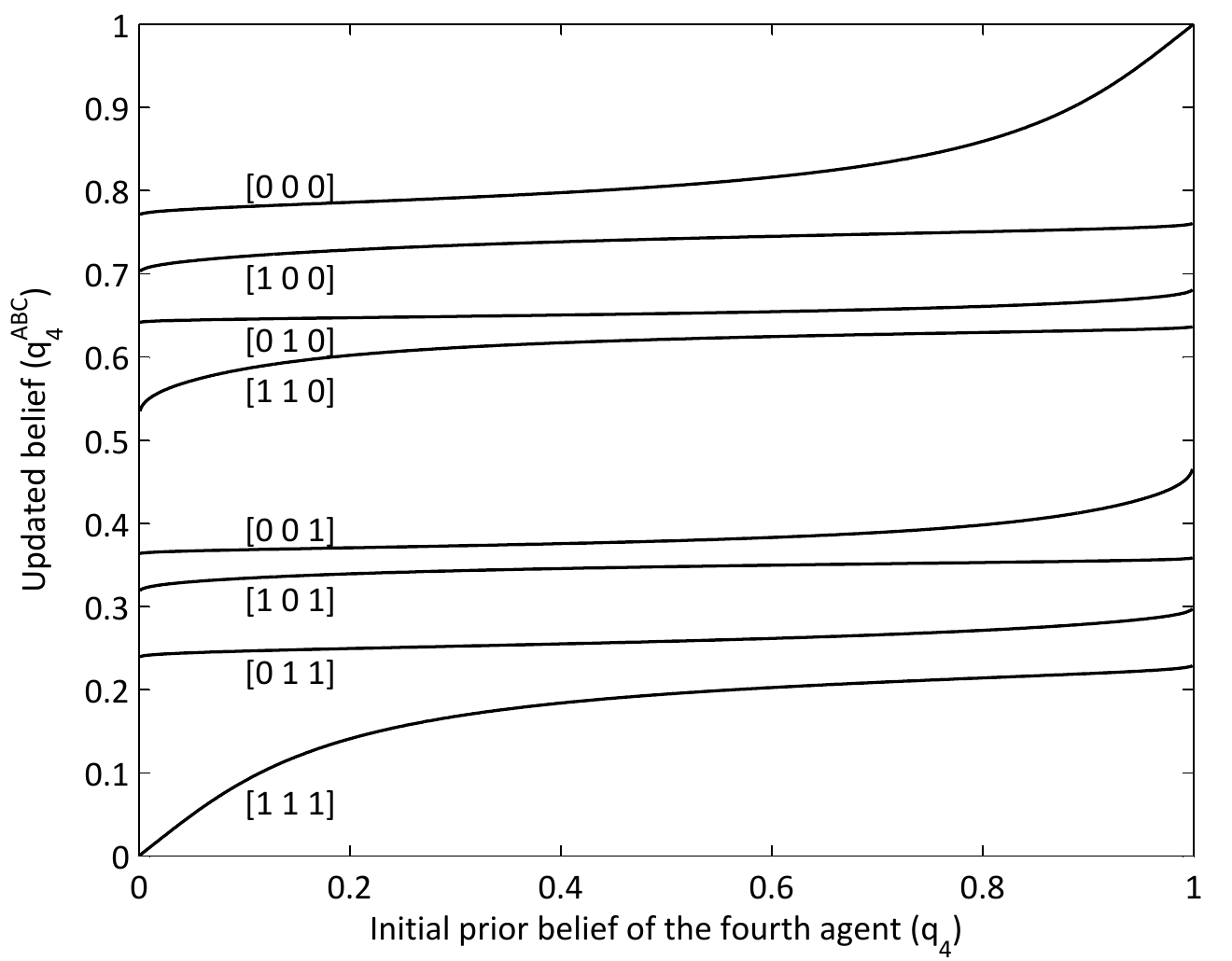}
    \caption{The function $U_4(q_4, \hhat_1, \hhat_2, \hhat_3)$---updated belief of the fourth agent ($\Belief_4^{_\text{ABC}}$)---for each possible combination of Alexis's, Blake's, and Chuck's decisions $[\hhat_1 \ \hhat_2 \ \hhat_3]$ when $c_{10} = c_{01} = 1$ and private signals are distorted by additive Gaussian noise with zero mean and unit variance.  The updated belief is mostly dependent on Chuck's decision; the top four curves are for $\hhat_3 = 0$ and the bottom four curves are for $\hhat_3 = 1$.}
    \label{fig:UpdatedBelief}
\end{figure}

Fig.~\ref{fig:UpdatedBelief} depicts the function $U_4(q_4, \hhat_1, \hhat_2, \hhat_3)$ for $N = 4$ for eight possible combinations of Alexis's, Blake's, and Chuck's decisions $[\hhat_1 \ \hhat_2 \ \hhat_3]$. An interesting property of $U_n$ is that the updated belief is much more dependent on the most recent decision $\hhat_{n-1}$ than on the earlier decisions $\hhat_1, \ldots, \hhat_{n-2}$.
This is especially the case when the $(n-1)$st agent has not followed precedent.  This is because the $n$th agent rationally concludes that the $(n-1)$st agent observed strong evidence to justify a deviation from precedent.  For example, if the decision history of the first five agents is $[0\ 0\ 0\ 0\ 1]$ then the sixth agent takes the last decision 1 seriously even though four agents chose 0\@.
A reversal of an arbitrarily long precedent sequence may occur because we assume unbounded private signals; if private signals are bounded like in~\cite{Banerjee1992,BikhchandaniHW1992}, then the influence of precedent can reach a point where agents cannot receive a signal strong enough to justify a decision running counter to precedent.

\section{Optimal Initial Belief}    \label{sec:OptimalInitialBelief}

We have constructed the prior belief update and decision making model in Section~\ref{sec:BeliefUpdate}.  In this section, we want to investigate when the system can achieve the minimum Bayes risk.  For simplicity, we only consider $N = 2$.  Note that the Bayes risk of the system is the same as Blake's Bayes risk because his decision is adopted as the final decision.

Let us recapitulate the computation of Blake's Bayes risk.
Alexis chooses her decision threshold as $\DecThres{1}{}{} = \lambda(\Belief_1)$.
Her probabilities of errors are given by
\begin{align*}
    \LocalISeq{1}{}{} & = p_{\Hhat_1 \MID H}(1 \MID 0) = \int_{\DecThres{1}{}{}}^{\infty} f_{Y_1 \MID H}(y \MID 0) \, dy, \nonumber \\
    \LocalIISeq{1}{}{} & = p_{\Hhat_1 \MID H}(0 \MID 1) = \int_{-\infty}^{\DecThres{1}{}{}} f_{Y_1 \MID H}(y \MID 1) \, dy.
\end{align*}

Blake thinks that Alexis uses the decision threshold $\DecThres{1}{}{\text{B}} = \lambda(\Belief_2)$
and computes her probabilities of errors differently:
\begin{subequations}
    \label{eq:Perror1-2}
\begin{align}
    \LocalISeq{1}{}{\text{B}} = p_{\Hhat_1 \MID H}(1 \MID 0)_{_{\text{B}}} & = \int_{\DecThres{1}{}{\text{B}}}^{\infty} f_{Y_1 \MID H}(y \MID 0) \, dy, \\
    \LocalIISeq{1}{}{\text{B}} = p_{\Hhat_1 \MID H}(0 \MID 1)_{_{\text{B}}} & = \int_{-\infty}^{\DecThres{1}{}{\text{B}}} f_{Y_1 \MID H}(y \MID 1) \, dy.
\end{align}
\end{subequations}

When Alexis decides $\Hhat_1 = 0$, Blake updates his belief $\Belief_2$ to $\Belief_2^{_0}$:
\begin{eqnarray}
    \lefteqn{\frac{\Belief_2^{_0}}{1 - \Belief_2^{_0}} = \frac{\Belief_2}{1 - \Belief_2} \frac{1 - \LocalISeq{1}{}{\text{B}}}{\LocalIISeq{1}{}{\text{B}}}} \nonumber \\
    & &\Longrightarrow
    \Belief_2^{_0} = \frac{\Belief_2 (1 - \LocalISeq{1}{}{\text{B}})}{\Belief_2 (1 - \LocalISeq{1}{}{\text{B}}) + (1 - \Belief_2) \LocalIISeq{1}{}{\text{B}}},
    \label{eq:BeliefUpdate2-0}
\end{eqnarray}
and his decision threshold is $\DecThres{2}{0}{} = \lambda(\Belief_2^{_0})$.
His probabilities of errors are given by
\begin{subequations}
    \label{eq:AlexisErrorProb-B0}
\begin{align}
    \LocalISeq{2}{0}{} & = p_{\Hhat_2 \MID \Hhat_1, H}(1 \MID 0, 0) = \int_{\DecThres{2}{0}{}}^{\infty} f_{Y_2 \MID H}(y \MID 0) \, dy, \\
    \LocalIISeq{2}{0}{} & = p_{\Hhat_2 \MID \Hhat_1, H}(0 \MID 0, 1) = \int_{-\infty}^{\DecThres{2}{0}{}} f_{Y_2 \MID H}(y \MID 1) \, dy.
\end{align}
\end{subequations}

Likewise, when Alexis decides $\Hhat_1 = 1$, Blake updates his belief $\Belief_2$ to $\Belief_2^{_1}$:
\begin{eqnarray}
    \lefteqn{\frac{\Belief_2^{_1}}{1 - \Belief_2^{_1}} = \frac{\Belief_2}{1 - \Belief_2} \frac{\LocalISeq{1}{}{\text{B}}}{1 - \LocalIISeq{1}{}{\text{B}}}} \nonumber \\
    & & \Longrightarrow
    \Belief_2^{_1} = \frac{\Belief_2 \LocalISeq{1}{}{\text{B}}}{\Belief_2 \LocalISeq{1}{}{\text{B}} + (1 - \Belief_2) (1 - \LocalIISeq{1}{}{\text{B}})},
    \label{eq:BeliefUpdate2-2}
\end{eqnarray}
and his decision threshold is $\DecThres{2}{1}{} = \lambda(\Belief_2^{_1})$.
His probabilities of errors are given by
\begin{subequations}
    \label{eq:AlexisErrorProb-B1}
\begin{align}
    \LocalISeq{2}{1}{} & = p_{\Hhat_2 \MID \Hhat_1, H}(1 \MID 1, 0) = \int_{\DecThres{2}{1}{}}^{\infty} f_{Y_2 \MID H}(y \MID 0) \, dy, \\
    \LocalIISeq{2}{1}{} & = p_{\Hhat_2 \MID \Hhat_1, H}(0 \MID 1, 1) = \int_{-\infty}^{\DecThres{2}{1}{}} f_{Y_2 \MID H}(y \MID 1) \, dy.
\end{align}
\end{subequations}

Now we can compute Blake's Bayes risk $R_2$:
\begin{align}
    R_2 & = c_{10} p_{\Hhat_2, H}(1, 0) + c_{01} p_{\Hhat_2, H}(0, 1) \nonumber \\
    & = c_{10} p_{\Hhat_2 \MID \Hhat_1, H}(1 \MID 0, 0) p_{\Hhat_1 \MID H}(0 \MID 0) p_{H}(0) \nonumber \\
    & \quad + c_{10} p_{\Hhat_2 \MID \Hhat_1, H}(1 \MID 1, 0) p_{\Hhat_1 \MID H}(1 \MID 0) p_{H}(0) \nonumber \\
    & \quad + c_{01} p_{\Hhat_2 \MID \Hhat_1, H}(0 \MID 0, 1) p_{\Hhat_1 \MID H}(0 \MID 1) p_{H}(1) \nonumber \\
    & \quad + c_{01} p_{\Hhat_2 \MID \Hhat_1, H}(0 \MID 1, 1) p_{\Hhat_1 \MID H}(1 \MID 1) p_{H}(1) \nonumber \\
    & = c_{10} \left[ \LocalISeq{2}{0}{} \, (1 - \LocalISeq{1}{}{}) + \LocalISeq{2}{1}{} \, \LocalISeq{1}{}{} \right] p_0 \nonumber \\
    & \quad + c_{01} \left[ \LocalIISeq{2}{0}{} \, \LocalIISeq{1}{}{} + \LocalIISeq{2}{1}{} \, (1 - \LocalIISeq{1}{}{}) \right] (1 - p_0).
    \label{eq:BlakeErrorProbability}
\end{align}

The Bayes risk $R_2$ in \eqref{eq:BlakeErrorProbability} is a function of $\Belief_1$ and $\Belief_2$.  It seems natural that $R_2$ is minimum at $\Belief_1 = \Belief_2 = p_0$ because Alexis will make the best decision she can and Blake will not misunderstand her decision.  Surprisingly, however, this turns out not to be true.  We will prove it by studying Alexis's optimal belief $\Belief_1^{\ast}$ with respect to minimizing $R_2$.

Let us consider the first derivative of \eqref{eq:BlakeErrorProbability}
with respect to $\Belief_1$:
\begin{equation*}
    \frac{d R_2}{d \Belief_1} = c_{10} p_0 (\LocalISeq{2}{1}{} - \LocalISeq{2}{0}{}) \frac{d \LocalISeq{1}{}{}}{d q_1} + c_{01} (1 - p_0) (\LocalIISeq{2}{0}{} - \LocalIISeq{2}{1}{}) \frac{d \LocalIISeq{1}{}{}}{d \Belief_1}.
\end{equation*}
We want to find $\Belief_1$ that makes this first derivative zero.
Using
\begin{align*}
    \frac{d \LocalISeq{1}{}{}}{d \Belief_1} & = \frac{d \LocalISeq{1}{}{}}{d \DecThres{1}{}{}} \frac{d \DecThres{1}{}{}}{d \Belief_1} = - f_{Y_1 \MID H}(\DecThres{1}{}{} \MID 0) \frac{d \DecThres{1}{}{}}{d \Belief_1}, \nonumber \\
    \frac{d \LocalIISeq{1}{}{}}{d \Belief_1} & = \frac{d \LocalIISeq{1}{}{}}{d \DecThres{1}{}{}} \frac{d \DecThres{1}{}{}}{d \Belief_1} = f_{Y_1 \MID H}(\DecThres{1}{}{} \MID 1) \frac{d \DecThres{1}{}{}}{d \Belief_1},
\end{align*}
this occurs when
\begin{eqnarray}
    \lefteqn{c_{10} p_0 (\LocalISeq{2}{1}{} - \LocalISeq{2}{0}{}) f_{Y_1 \MID H}(\DecThres{1}{}{} \MID 0)} \nonumber \\
    & & \qquad = c_{01} (1 - p_0) (\LocalIISeq{2}{0}{} - \LocalIISeq{2}{1}{}) f_{Y_1 \MID H}(\DecThres{1}{}{} \MID 1) \nonumber \\
    \lefteqn{\Longleftrightarrow \frac{f_{Y_1 \MID H}(\DecThres{1}{}{} \MID 1)}{f_{Y_1 \MID H}(\DecThres{1}{}{} \MID 0)} = \frac{c_{10} p_0 (\LocalISeq{2}{1}{} - \LocalISeq{2}{0}{})}{c_{01} (1 - p_0) (\LocalIISeq{2}{0}{} - \LocalIISeq{2}{1}{})}.}
    \label{eq:Derivative1}
\end{eqnarray}
Note that $\DecThres{1}{}{} = \lambda(\Belief_1)$ is a solution to \eqref{eq:Threshold},
\begin{equation*}
    \frac{f_{Y_1 \MID H} (\DecThres{1}{}{} \MID 1)}{f_{Y_1 \MID H} (\DecThres{1}{}{} \MID 0)} = \frac{c_{10} \Belief_1}{c_{01} (1 - \Belief_1)}.
\end{equation*}
Therefore Alexis's optimal belief $\Belief_1^{\ast}$ needs to satisfy
\begin{equation}
    \frac{\Belief_1^{\ast}}{1 - \Belief_1^{\ast}} =\frac{p_0 (\LocalISeq{2}{1}{} - \LocalISeq{2}{0}{})}{(1 - p_0) (\LocalIISeq{2}{0}{} - \LocalIISeq{2}{1}{})},
    \label{eq:Derivative1-1}
\end{equation}
where the value of $\pFrac{\LocalISeq{2}{1}{} - \LocalISeq{2}{0}{}}{\LocalIISeq{2}{0}{} - \LocalIISeq{2}{1}{}}$ does not have to be 1\@.  In fact, in additive Gaussian noise cases, it is not equal to one except for $p_0 = \Frac{c_{01}}{(c_{10} + c_{01})}$.  Therefore, the optimal value of $\Belief_1$ is not $p_0$ in general.

\section{Example: Gaussian Likelihoods}    \label{sec:Example}

Suppose that the $n$th agent receives the signal $Y_n = H + W_n$ where the additive noises $W_n$ are iid with pdf
\begin{equation}
    f_W(w) = \frac{1}{\sqrt{2 \pi}} e^{-w^2 / 2}.
    \label{eq:NoisePDF}
\end{equation}
Each likelihood $f_{Y_n|H}$ is thus Gaussian with mean $H$ and variance $\sigma^2$.  For a prior belief $\Belief_n$, the likelihood ratio test
\begin{equation}
    \frac{f_{Y_n \MID H} (y_n \MID 1)}{f_{Y_n \MID H} (y_n \MID 0)} \overset{\Hhat_n(y_n) = 1}{\underset{\Hhat_n(y_n) = 0}{\gtreqless}} \frac{c_{10} \Belief_n}{c_{01} (1 - \Belief_n)}
\end{equation}
is simplified to the following comparison:
\begin{equation}
    y_n \overset{\Hhat_n(y_n) = 1}{\underset{\Hhat_n(y_n) = 0}{\gtreqless}} \DecThres{n}{}{}  = \frac{1}{2} + \log \frac{c_{10} \Belief_n}{c_{01} (1 - \Belief_n)}.
    \label{eq:ThresholdGaussian}
\end{equation}

\begin{figure}
    \centering
    \includegraphics[width=3.2in]{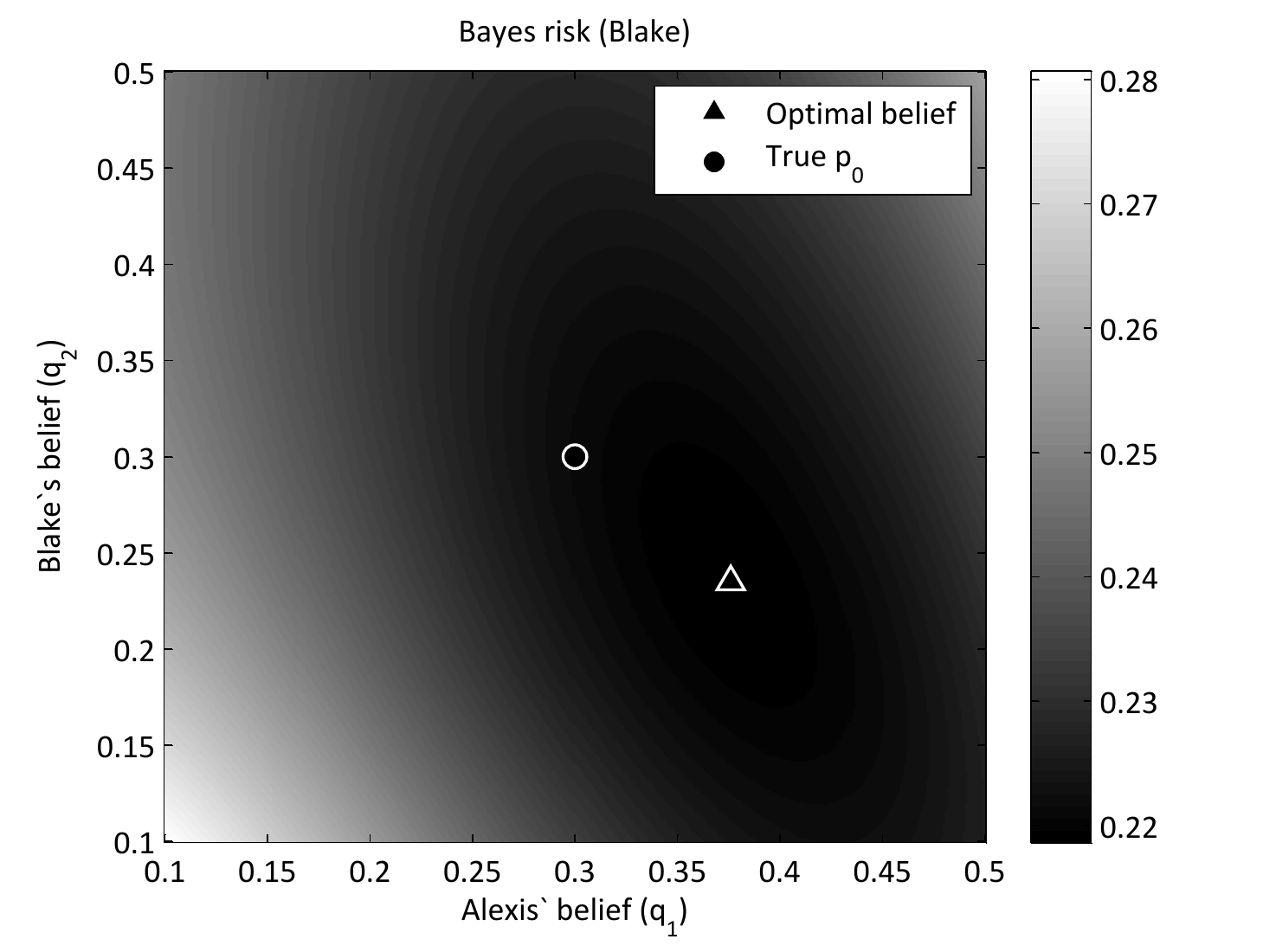}
    \caption{Visualization of the Bayes risk for various $q_1$ and $q_2$ for $N = 2$, $c_{10} = c_{01} = 1$, $p = 0.3$, and additive Gaussian noise with zero mean and unit variance.  Alexis's and Blake's optimal prior beliefs ($\filledmedtriangleup$) are different from the true prior probability ($\bullet$).}
    \label{fig:BRContour1}
\end{figure}

Fig.~\ref{fig:BRContour1} clearly shows that knowing true prior probability is not optimal.  We have computed the performance of the sequential decision making by two agents for $c_{10} = c_{01} = 1$, $p = 0.3$, and additive Gaussian noise with zero mean and unit variance.
The Bayes risk is minimum when Alexis perceives the prior probability as 0.38 and Blake perceives it as 0.23, which is marked with a triangle.  For convenience, the true probability is indicated by a circle.

\begin{figure}
    \centering{
    \subfloat[]{\includegraphics[width=1.8in]{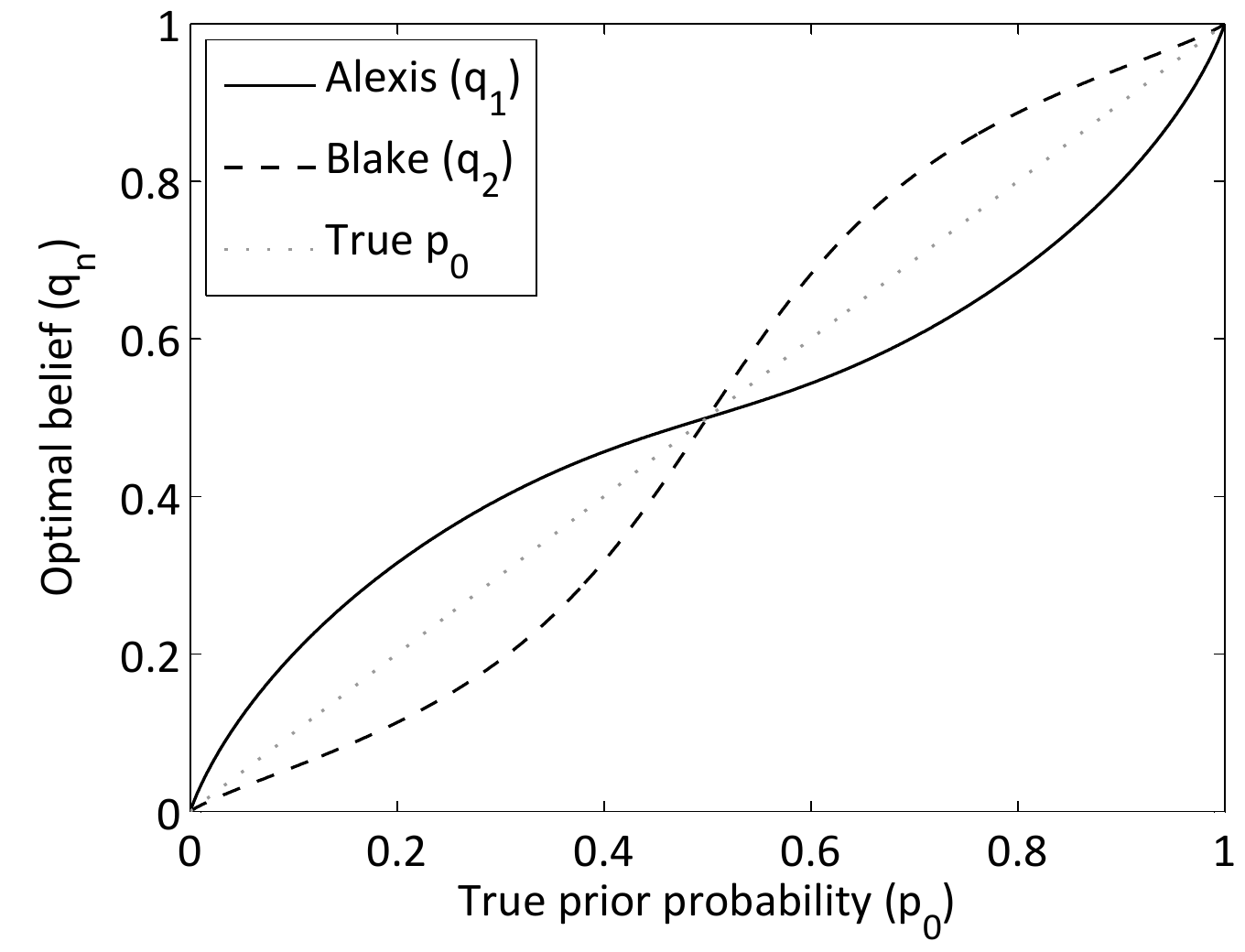}}
    \subfloat[]{\includegraphics[width=1.8in]{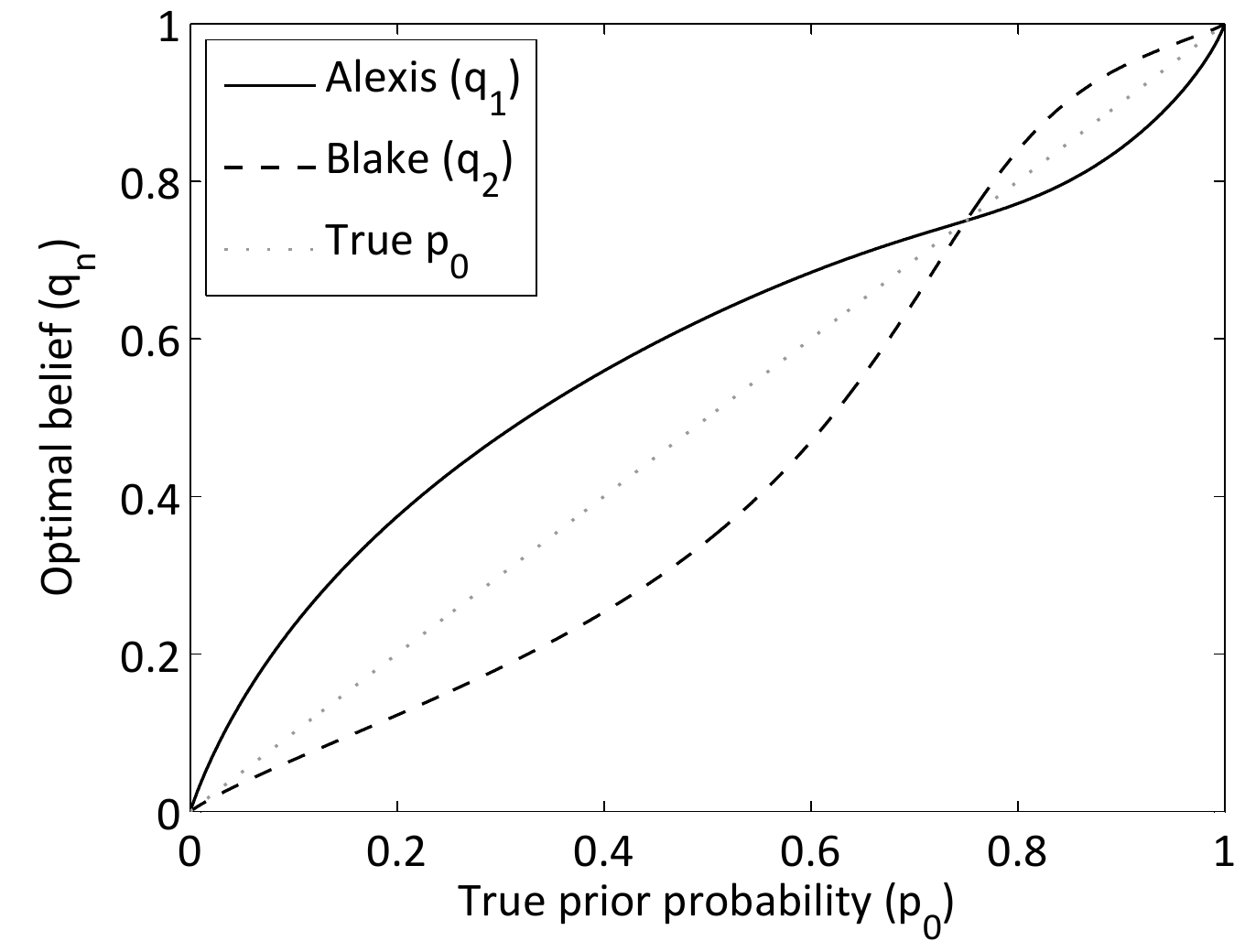}}
    }
    \caption{The trend of the optimal prior beliefs for varying $p_0$ for $N = 2$ (Alexis and Blake).  (a) $c_{10} = c_{01} = 1$.  (b) $c_{10} = 1$, $c_{01} = 3$.}
    \label{fig:Trend1}
\end{figure}

\begin{figure}
    \centering{
    \subfloat[]{\includegraphics[width=1.8in]{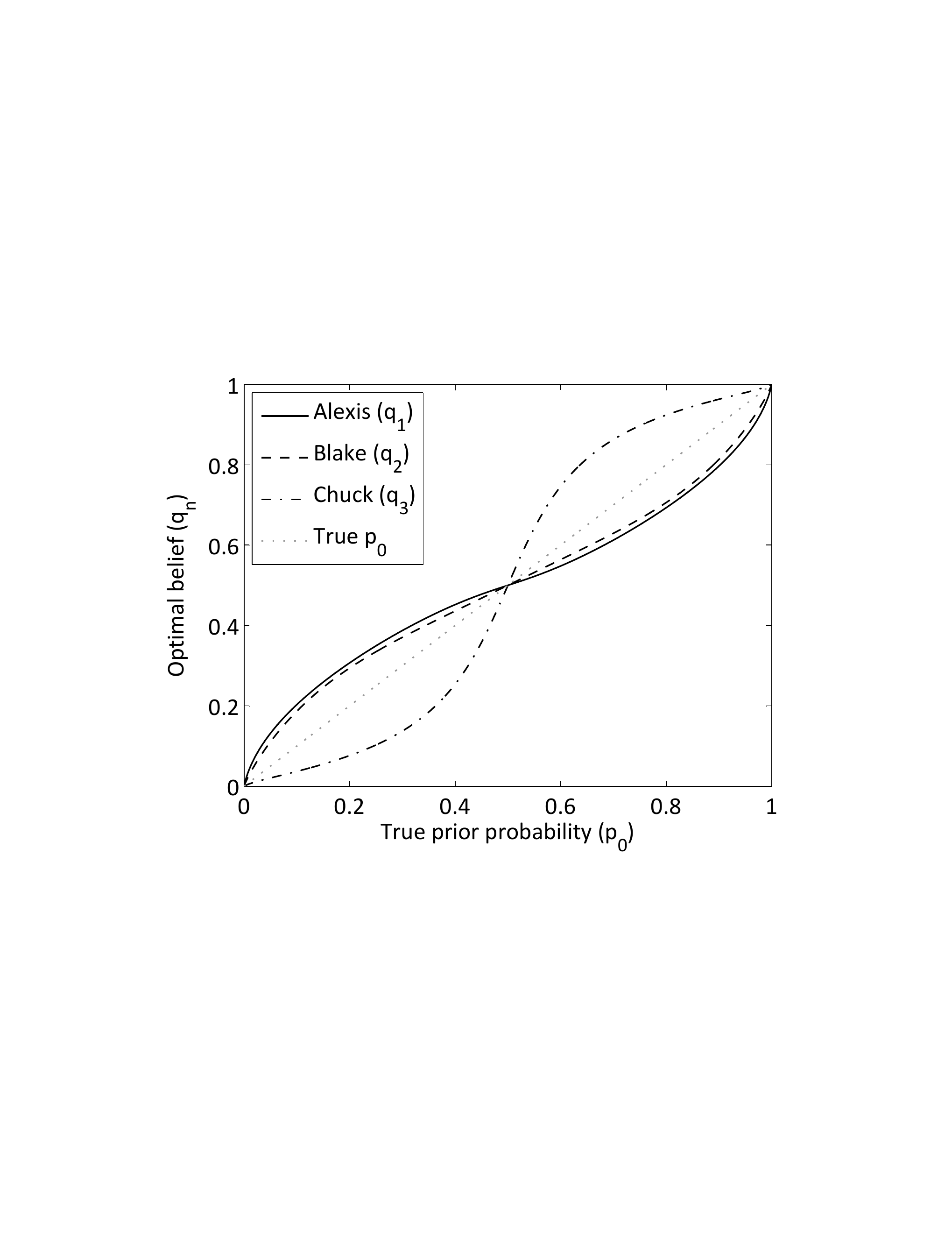}}
    \subfloat[]{\includegraphics[width=1.8in]{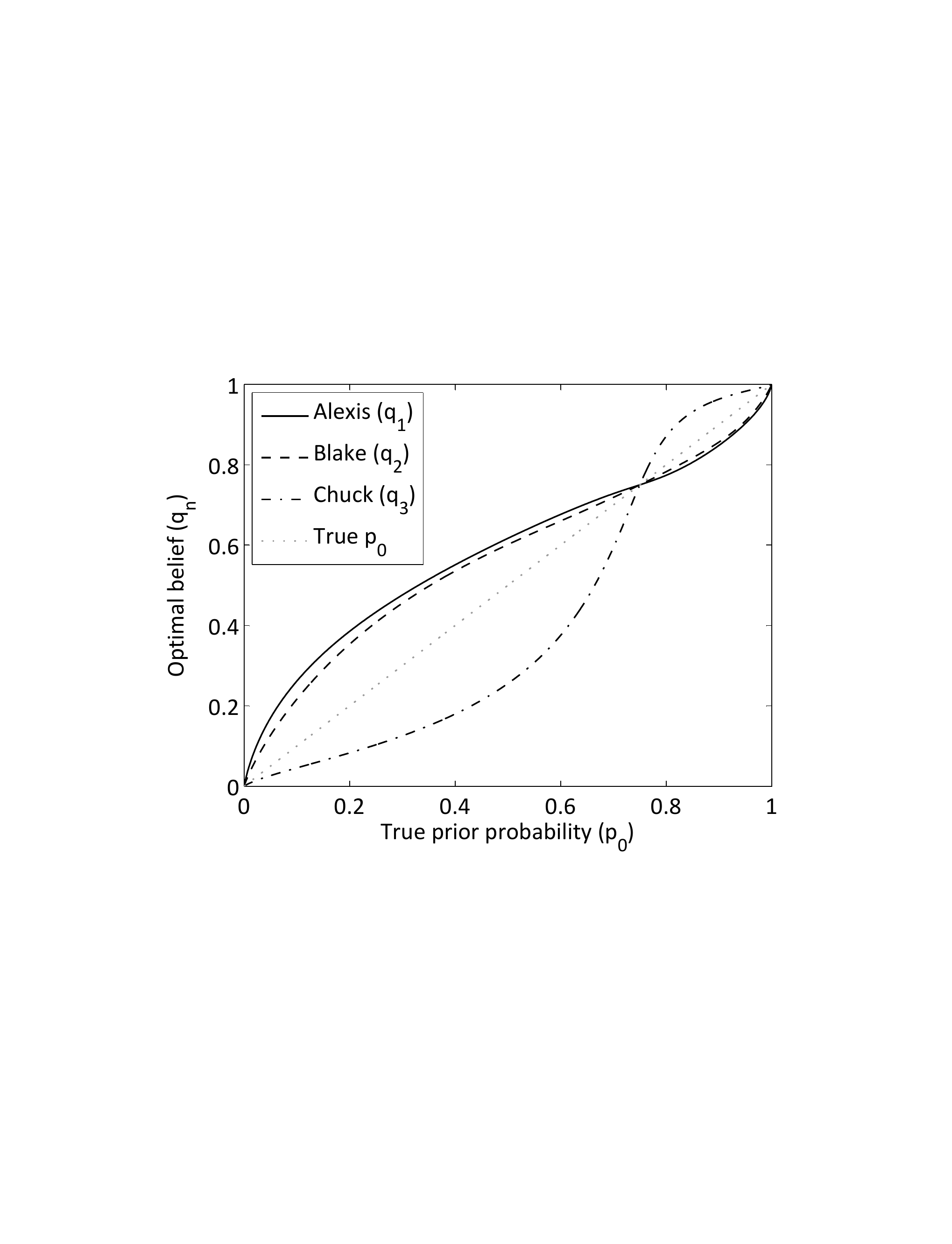}}
    }
    \caption{The trend of the optimal prior beliefs for varying $p_0$ for $N = 3 $ (Alexis, Blake, and Chuck).  (a) $c_{10} = c_{01} = 1$.  (b) $c_{10} = 1$, $c_{01} = 3$.}
    \label{fig:Trend2}
\end{figure}

Figs.~\ref{fig:Trend1} and~\ref{fig:Trend2} depict the prior beliefs that the agents should have for optimal decision making.  They show several common characteristics for the additive Gaussian noise model:  First, the non-terminal agents (i.e., Alexis for $N=2$ and Alexis and Blake for $N=3$) should have belief larger than $p_0$ when $p_0$ is small and belief smaller than $p_0$ when $p_0$ is large.  We call this \emph{open-mindedness} because it is to assign higher prior belief to outcomes that are very unlikely.  Second, the last agent (i.e., Blake for $N = 2$ and Chuck for $N = 3$) should have belief smaller than $p_0$ when $p_0$ is small and belief larger than $p_0$ when $p_0$ is large.  This is necessary to compensate for the biases of precedent agents.  Last, there is a unique point, except for $p_0 = 0$ or $p_0 = 1$, where all agents' optimal prior beliefs are the same as the true prior probability.  It occurs at $p_0 = c_{01} / (c_{10} + c_{01})$.  We prove this for $N = 2$ in the appendix.  We also show there for $N = 2$ that the first agent should be open minded.

\section{Conclusion}    \label{sec:Conclusion}
We have discussed decision making sequentially performed by a group of agents that make decisions based on individually biased prior beliefs.  Instead of investigating herding on a wrong action, we have assumed unbounded private signals and focused on the agents' belief update.  The Bayes-optimal updated belief turns out to be the probability of each hypothesis conditioned on the decisions made by previous agents.  It gets more difficult for agents in later positions to make decisions that differ from precedents; however, if one observes a very strong signal against the precedents and chooses the alternative, the decision will be taken very seriously by the following agents.

The wrong beliefs held by previous agents change the probability that following agents choose each hypothesis.  Contrary to intuition, however, wrong beliefs are not always bad.  In fact, the optimal beliefs of agents (those that lead to the minimum Bayes risk for the last agent) are not usually equal to the true prior probability.  Specifically, in the case of observations corrupted by iid additive Gaussian noises, an agent biased toward $c_{01} / (c_{10} + c_{01})$ can be more beneficial to subsequent agents than is an accurate agent is.  The point $c_{01} / (c_{10} + c_{01})$ is special because the probabilities of false alarms and missed detections will be balanced by the optimal decision rule at the prior probability.  In terms of human decision making, where precedent agents are advisers or counselors to the last agent who has the final decisive power, we can say that the best advisers are necessarily open-minded people.

The idea of an open-minded adviser is related to the amount of information conveyed in the public signals.  A public signal is a quantized version of a private signal while also simultaneously reflecting an agent's belief.  Alexis's decision will reflect her belief more than her private signal when her belief is very small (close to 0) or very large (close to 1).  However, Blake would want a public signal that is most informative of Alexis's private signal.  Therefore, he wants her to make her decision with a less extreme belief or an open mind.

While some conclusions of our study depend on having Gaussian likelihoods and may not hold for different types of additive noise, it is more generally true that the optimal prior beliefs are different from the true prior probability.


\appendix[Alexis's Optimal Prior Belief for Blake]
For the case of $N = 2$, we investigate Alexis's prior belief that minimizes Blake's Bayes risk.  Let us assume that
\begin{subequations}
    \label{eq:NoisePDF_h1&sigma}
\begin{align}
    f_{Y \MID H}(y \MID 0) & = \frac{1}{\sqrt{2 \pi \sigma^2}} \exp\left({-\frac{y^2}{2 \sigma^2}}\right), \\
    f_{Y \MID H}(y \MID 1) & = \frac{1}{\sqrt{2 \pi \sigma^2}} \exp\left({-\frac{(y - h_1)^2}{2 \sigma^2}}\right),
\end{align}
\end{subequations}
where $\sigma$ and $h_1$ are arbitrary positive numbers.

\begin{figure}
    \centering{
    \subfloat[]{\includegraphics[width=1.8in]{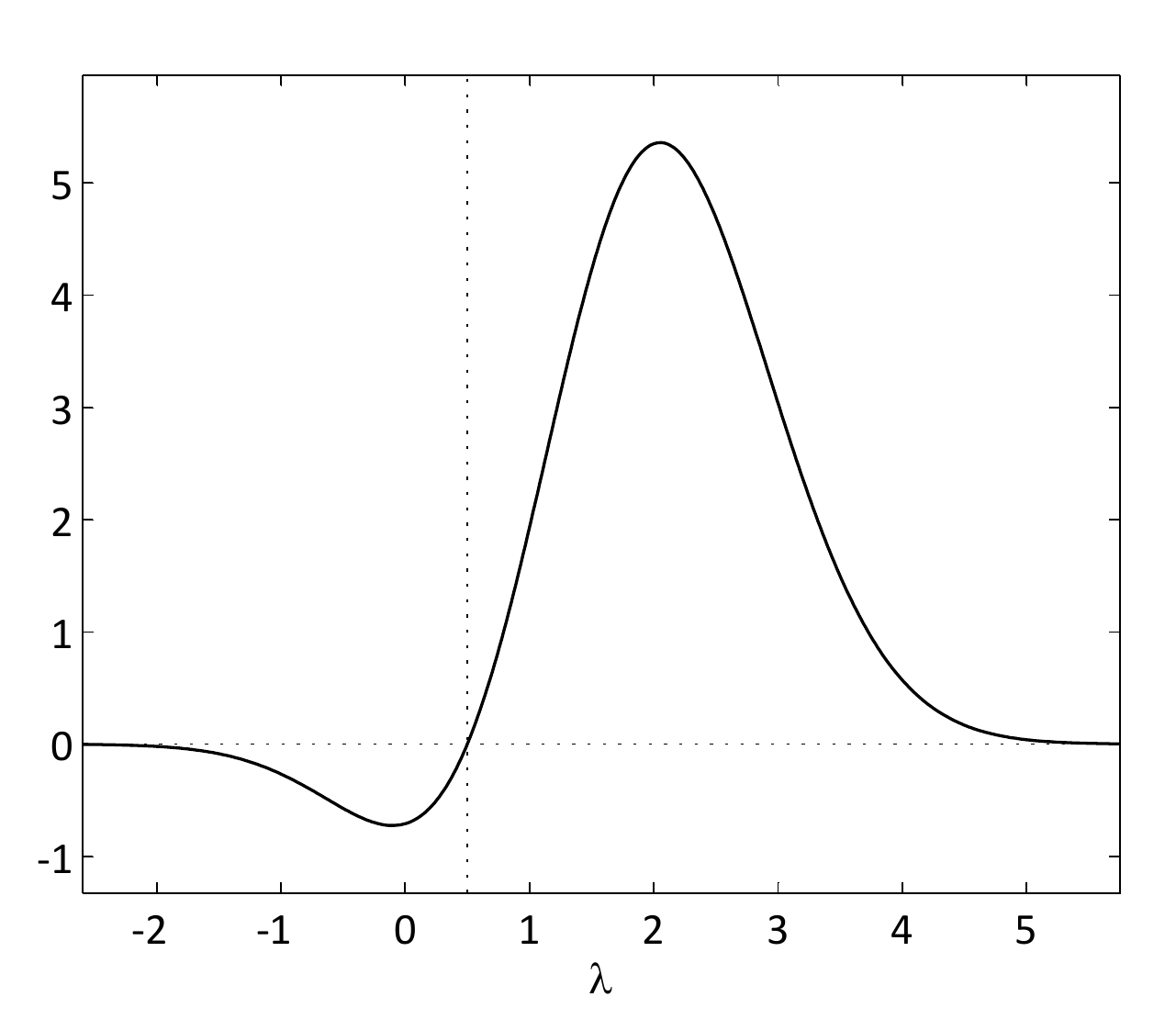}}
    \subfloat[]{\includegraphics[width=1.8in]{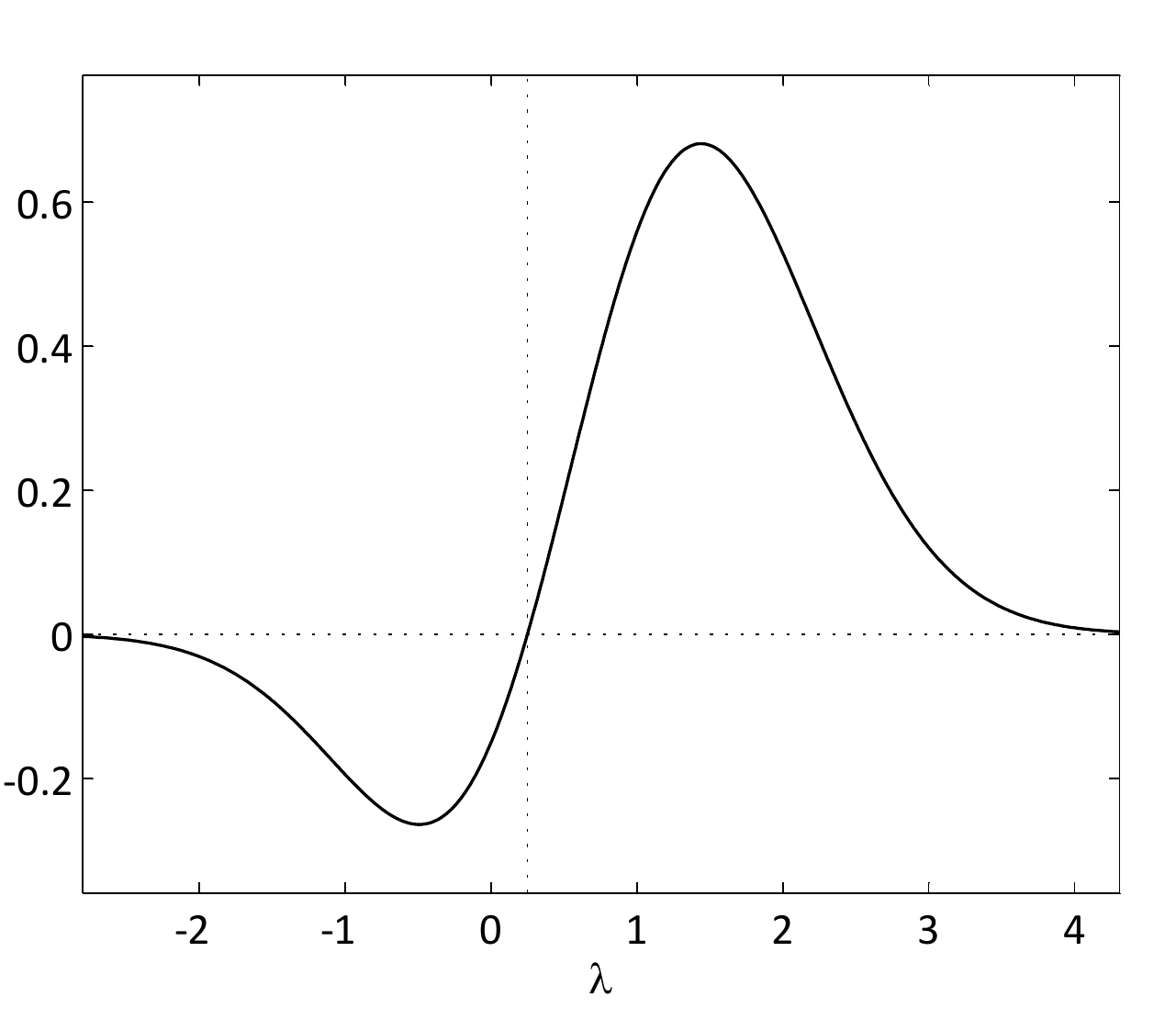}}
    }
    \caption{The difference between the left-hand side and the right-hand side of \eqref{eq:BasicInequality1}.  (a) $h_1 = 1$.  (b) $h_1 = 0.5$.}
    \label{fig:Conjecture1}
\end{figure}

\begin{conjecture}
    If $\lambda < \Frac{h_1}{2}$, then
    \begin{align}
        & \int_{-\infty}^{\lambda} \exp\left(-\frac{y^2}{2} + \lambda h_1\right) \, dy \int_{\lambda}^{\infty} \exp\left(-\frac{y^2}{2} + \lambda h_1\right) \, dy \nonumber \\
        & < \int_{-\infty}^{\lambda} \exp\left(-\frac{y^2}{2} + y h_1\right) \, dy \int_{\lambda}^{\infty} \exp\left(-\frac{y^2}{2} + y h_1\right) \, dy.
        \label{eq:BasicInequality1}
    \end{align}
    \label{conj:BasicInequaility1}
\end{conjecture}
Fig.~\ref{fig:Conjecture1} depicts the difference between the left-hand side and the right-hand side of \eqref{eq:BasicInequality1} for two values of $h_1$ and supports the conjecture.
In the following we assume the conjecture to be true.

\begin{lemma}
    If $\lambda < \Frac{h_1}{2}$, then
    \begin{align}
        & \int_{-\infty}^{\lambda} \exp\left(-\frac{y^2}{2 \sigma^2} + \frac{\lambda h_1}{\sigma^2}\right) \, dy \int_{\lambda}^{\infty} \exp\left(-\frac{y^2}{2 \sigma^2} + \frac{\lambda h_1}{\sigma^2}\right) \, dy \nonumber \\
        & < \int_{-\infty}^{\lambda} \exp\left(-\frac{y^2}{2 \sigma^2} + \frac{y h_1}{\sigma^2}\right) \, dy \int_{\lambda}^{\infty} \exp\left(-\frac{y^2}{2} + \frac{y h_1}{\sigma^2}\right) \, dy.
        \label{eq:BasicInequality2}
    \end{align}
\end{lemma}
\begin{IEEEproof}
    Substituting $y' = y / \sigma$, $\lambda' = \lambda / \sigma$, and $h_1' = h_1' / \sigma$, we obtain
    \begin{align*}
        & \int_{-\infty}^{\lambda} \exp\left(-\frac{y^2}{2 \sigma^2} + \frac{\lambda h_1}{\sigma^2}\right) \, dy \int_{\lambda}^{\infty} \exp\left(-\frac{y^2}{2 \sigma^2} + \frac{\lambda h_1}{\sigma^2}\right) \, dy \nonumber \\
        & = \sigma^2 \int_{-\infty}^{\lambda'} \exp\left(-\frac{y'^2}{2} + \lambda' h_1'\right) \, dy' \int_{\lambda'}^{\infty} \exp\left(-\frac{y'^2}{2} + \lambda' h_1'\right) \, dy'
    \end{align*}
    and
    \begin{align*}
        & \int_{-\infty}^{\lambda} \exp\left(-\frac{y^2}{2 \sigma^2} + \frac{y h_1}{\sigma^2}\right) \, dy \int_{\lambda}^{\infty} \exp\left(-\frac{y^2}{2} + \frac{y h_1}{\sigma^2}\right) \, dy \nonumber \\
        & = \sigma^2 \int_{-\infty}^{\lambda'} \exp\left(-\frac{y'^2}{2} + y' h'_1\right) \, dy' \int_{\lambda'}^{\infty} \exp\left(-\frac{y'^2}{2} + y' h'_1\right) \, dy'.
    \end{align*}
    Since $\lambda < \Frac{h_1}{2}$ implies $\lambda' < \Frac{h'_1}{2}$, \eqref{eq:BasicInequality2} follows from Conjecture~\ref{conj:BasicInequaility1}.
\end{IEEEproof}

\begin{theorem}
    Alexis's and Blake's optimal prior beliefs are equal to the true prior probability $p_0$ if $p_0 = \Frac{c_{01}}{(c_{10} + c_{01})}$.
    \label{thm:TrueIsOptimal}
\end{theorem}
\begin{IEEEproof}
    We will show that $\Frac{\partial R_2}{\partial q_1} = 0$ and $\Frac{\partial R_2}{\partial q_2} = 0$ for $q_1 = q_2 = p_0 = \Frac{c_{01}}{(c_{10} + c_{01})}$.  Then they are Alexis's and Blake's optimal prior beliefs $\Belief_1^{\ast}$ and $\Belief_2^{\ast}$.

    First, consider $\Frac{\partial R_2}{\partial q_2}$ using~\eqref{eq:BlakeErrorProbability}:
    {\footnotesize
    \begin{align}
        &\frac{\partial R_2}{\partial q_2} \nonumber \\
        & = - c_{10} p_0 \left[ (1 - \LocalISeq{1}{}{}) f_{Y_2 \MID H}(\DecThres{2}{0}{} \MID 0) \frac{d \DecThres{2}{0}{}}{d q_2} + \LocalISeq{1}{}{} f_{Y_2 \MID H}(\DecThres{2}{1}{} \MID 0) \frac{d \DecThres{2}{1}{}}{d q_2} \right] \nonumber \\
        & \quad + c_{01} (1 - p_0) \left[ \LocalIISeq{1}{}{} f_{Y_2 \MID H}(\DecThres{2}{0}{} \MID 1) \frac{d \DecThres{2}{0}{}}{d q_2} + (1 - \LocalIISeq{1}{}{}) f_{Y_2 \MID H}(\DecThres{2}{1}{} \MID 1) \frac{d \DecThres{2}{1}{}}{d q_2} \right].
        \label{eq:Derivative2}
    \end{align}
    }From~\eqref{eq:BeliefUpdate2-0} and~\eqref{eq:ThresholdGaussian},
\begin{displaymath}
\DecThres{2}{0}{} = \DecThres{1}{}{\text{B}} + \log \frac{1 - \LocalISeq{1}{}{\text{B}}}{\LocalIISeq{1}{}{\text{B}}},
\end{displaymath}
and its derivative is given by
    \begin{align}
        \frac{d \DecThres{2}{0}{}}{d q_2} & = \frac{d \DecThres{1}{}{\text{B}}}{d q_2} - \frac{d \LocalISeq{1}{}{\text{B}}}{d q_2} \frac{1}{1 - \LocalISeq{1}{}{\text{B}}} - \frac{d \LocalIISeq{1}{}{\text{B}}}{d q_2} \frac{1}{\LocalIISeq{1}{}{\text{B}}} \nonumber \\
        & = \left[ 1 + \frac{f_{Y_1 \MID H}(\DecThres{1}{}{\text{B}} \MID 0)}{1 - \LocalISeq{1}{}{\text{B}}} - \frac{f_{Y_1 \MID H}(\DecThres{1}{}{\text{B}} \MID 1)}{\LocalIISeq{1}{}{\text{B}}} \right] \frac{d \DecThres{1}{}{\text{B}}}{d q_2}.
        \label{eq:Derivative2-0}
    \end{align}
    Likewise,
\begin{displaymath}
\DecThres{2}{1}{} = \DecThres{1}{}{\text{B}} + \log \frac{\LocalISeq{1}{}{\text{B}}}{1 - \LocalIISeq{1}{}{\text{B}}},
\end{displaymath}
and its derivative is
    \begin{align}
        \frac{d \DecThres{2}{1}{}}{d q_2} = \left[ 1 - \frac{f_{Y_1 \MID H}(\DecThres{1}{}{\text{B}} \MID 0)}{\LocalISeq{1}{}{\text{B}}} + \frac{f_{Y_1 \MID H}(\DecThres{1}{}{\text{B}} \MID 1)}{1 - \LocalIISeq{1}{}{\text{B}}} \right] \frac{d \DecThres{1}{}{\text{B}}}{d q_2}.
        \label{eq:Derivative2-1}
    \end{align}

    In addition, $q_1 = q_2$ implies that $\LocalISeq{1}{}{\text{B}} = \LocalISeq{1}{}{}$ and $\LocalIISeq{1}{}{\text{B}} = \LocalIISeq{1}{}{}$, and we can derive the following relations for $q_1 = q_2 = p_0$:
    \begin{align}
        \frac{f_{Y_2 \MID H} (\DecThres{2}{0}{} \MID 1)}{f_{Y_2 \MID H} (\DecThres{2}{0}{} \MID 0)} & = \frac{c_{10} \Belief_2 (1 - \LocalISeq{1}{}{\text{B}})}{c_{01}(1 - \Belief_2) \LocalIISeq{1}{}{\text{B}}} = \frac{c_{10} p_0 (1 - \LocalISeq{1}{}{})}{c_{01}(1 - p_0) \LocalIISeq{1}{}{}}, \nonumber \\
        \frac{f_{Y_2 \MID H} (\DecThres{2}{1}{} \MID 1)}{f_{Y_2 \MID H} (\DecThres{2}{1}{} \MID 0)} & = \frac{c_{10} \Belief_2 \LocalISeq{1}{}{\text{B}}}{c_{01}(1 - \Belief_2)(1 - \LocalIISeq{1}{}{\text{B}})} = \frac{c_{10} p_0 \LocalISeq{1}{}{}}{c_{01}(1 - p_0) (1 - \LocalIISeq{1}{}{})}.
        \label{eq:Relations2}
    \end{align}

    By substituting~\eqref{eq:Derivative2-0} and~\eqref{eq:Derivative2-1} in~\eqref{eq:Derivative2} and using the relations~\eqref{eq:Relations2}, we obtain that $\Frac{\partial R_2}{\partial q_2} = 0$ at $q_1 = q_2 = p_0$.

    Next, we consider $\Frac{\partial R_2}{\partial q_1}$, which is zero at $q_1$ and $q_2$ that satisfy~\eqref{eq:Derivative1-1},
    \begin{equation*}
        \frac{q_1}{(1 - q_1)} = \frac{p_0 (\LocalISeq{2}{1}{} - \LocalISeq{2}{0}{})}{(1 - p_0) (\LocalIISeq{2}{0}{} - \LocalIISeq{2}{1}{})}.
    \end{equation*}
    The condition $q_2 = \Frac{c_{01}}{(c_{10} + c_{01})}$ leads to $\DecThres{1}{}{\text{B}} = \Frac{h_1}{2}$ and $\LocalISeq{1}{}{\text{B}} = \LocalIISeq{1}{}{\text{B}}$.  Hence, $\DecThres{2}{0}{} - \DecThres{1}{}{\text{B}} = \DecThres{1}{}{\text{B}} - \DecThres{2}{1}{}$ and $\DecThres{2}{0}{} = 1 - \DecThres{2}{1}{}$.  Then, from~\eqref{eq:AlexisErrorProb-B0} and~\eqref{eq:AlexisErrorProb-B1}, we obtain $\LocalISeq{2}{0}{} = \LocalIISeq{2}{1}{}$ and $\LocalISeq{2}{1}{} = \LocalIISeq{2}{0}{}$.  Therefore, only $q_2 = p_0$ completes~\eqref{eq:Derivative1-1} and makes $\Frac{\partial R_2}{\partial q_1}$ zero.
\end{IEEEproof}

\begin{theorem}
    Let $p_0 \in (0, 1)$ denote the true prior probability and $\Belief_1^{\ast}$ Alexis's (i.e., the first agent's) optimal prior belief.
\begin{itemize}
\item
 If $p_0 < \Frac{c_{01}}{(c_{10} + c_{01})}$, then $p_0 < \Belief_1^{\ast} < \Frac{c_{01}}{(c_{10} + c_{01})}$.
\item
 If $p_0 = \Frac{c_{01}}{(c_{10} + c_{01})}$, then $\Belief_1^{\ast} = p_0$.
\item
 If $p_0 > \Frac{c_{01}}{(c_{10} + c_{01})}$, then $\Frac{c_{01}}{(c_{10} + c_{01})} < \Belief_1^{\ast} < p_0$.
\end{itemize}
\end{theorem}
\begin{IEEEproof}
    First, the proof for the case when $p_0 = \Frac{c_{01}}{(c_{10} + c_{01})}$ is given in Theorem~\ref{thm:TrueIsOptimal}.

    Next, suppose that $p_0 < \Frac{c_{01}}{(c_{10} + c_{01})}$.  Let $\lambda$ in~\eqref{eq:BasicInequality2} denote a decision threshold according to $\Belief_2^{\ast}$.  Obviously, optimal prior beliefs $\Belief_1^{\ast}$ and $\Belief_2^{\ast}$ should be strictly decreasing as $p_0$ decreases like in Fig.~\ref{fig:UpdatedBelief}.  Hence, Theorem~\ref{thm:TrueIsOptimal}, which states that $\Belief_1^{\ast} = \Belief_2^{\ast} = \Frac{c_{01}}{(c_{10} + c_{01})}$ if $p_0 = \Frac{c_{01}}{(c_{10} + c_{01})}$, implies that
    \begin{equation}
        \Belief_1^{\ast} < \frac{c_{01}}{c_{10} + c_{01}} \quad \mbox{and} \quad \Belief_2^{\ast} < \frac{c_{01}}{c_{10} + c_{01}}
    \end{equation}
    if $p_0 < \Frac{c_{01}}{(c_{10} + c_{01})}$.  Then we get $\lambda < \Frac{h_1}{2}$ and can use~\eqref{eq:BasicInequality2}.

    Multiplying each integrand in~\eqref{eq:BasicInequality2} by the constant $\frac{1}{2 \pi \sigma^2} \exp(-\frac{\lambda^2}{2 \sigma^2} - \frac{h_1^2}{2 \sigma^2})$, we get
    \begin{align}
        & \int_{-\infty}^{\lambda} \frac{1}{2 \pi \sigma^2} \exp\left(-\frac{(\lambda - h_1)^2}{2 \sigma^2}\right) \exp\left(-\frac{y^2}{2 \sigma^2}\right) \, dy \nonumber \\
        & \times \int_{\lambda}^{\infty} \frac{1}{2 \pi \sigma^2} \exp\left(-\frac{(\lambda - h_1)^2}{2 \sigma^2}\right) \exp\left(-\frac{y^2}{2 \sigma^2}\right) \, dy \nonumber \\
        & < \int_{-\infty}^{\lambda} \frac{1}{2 \pi \sigma^2} \exp\left(-\frac{\lambda^2}{2 \sigma^2}\right) \exp\left(-\frac{(y - h_1)^2}{2 \sigma^2}\right) \, dy \nonumber \\
        & \quad \times \int_{\lambda}^{\infty} \frac{1}{2 \pi \sigma^2} \exp\left(-\frac{\lambda^2}{2 \sigma^2}\right) \exp\left(-\frac{(y - h_1)^2}{2 \sigma^2}\right) \, dy.
        \label{eq:BasicInequality3}
    \end{align}
    According to~\eqref{eq:NoisePDF_h1&sigma}, the exponential functions in~\eqref{eq:BasicInequality3} are likelihood functions of $Y_1$, so we have
    \begin{align}
        \lefteqn{f_{Y_1 \MID H}^2(\lambda \MID 1) \int_{-\infty}^{\lambda}  f_{Y_1 \MID H}(y \MID 0) \, dy
        \int_{\lambda}^{\infty} f_{Y_1 \MID H}(y \MID 0) \, dy} \nonumber \\
        & < f_{Y_1 \MID H}^2(\lambda \MID 0) \int_{-\infty}^{\lambda} f_{Y_1 \MID H}(y \MID 1) \, dy
        \int_{\lambda}^{\infty} f_{Y_1 \MID H}(y \MID 1) \, dy.
        \label{eq:BasicInequality4}
    \end{align}
    Since $\lambda = \lambda(q_2^{\ast})$, using~\eqref{eq:Perror1-2}, we obtain
    \begin{align}
        \frac{\LocalISeq{1}{}{\text{B}} (1 - \LocalISeq{1}{}{\text{B}})}{f_{Y_1 \MID H}^2(\lambda \MID 0)}
        < \frac{\LocalIISeq{1}{}{\text{B}} (1 - \LocalIISeq{1}{}{\text{B}})}{f_{Y_1 \MID H}^2(\lambda \MID 1)},
        \label{eq:Inequality4}
    \end{align}
    and \eqref{eq:Threshold} transforms \eqref{eq:Inequality4} to
    \begin{align}
        c_{10}^2 \Belief_2^{\ast 2} \LocalISeq{1}{}{\text{B}} (1 - \LocalISeq{1}{}{\text{B}})
        < c_{01}^2 (1 - \Belief_2^{\ast})^2 \LocalIISeq{1}{}{\text{B}} (1 - \LocalIISeq{1}{}{\text{B}}).
        \nonumber
    \end{align}
    Rearranging terms gives us
    \begin{align}
        \left( \frac{c_{10} q_2^{\ast}}{c_{01} (1 - q_2^{\ast})} \frac{1 - \LocalISeq{1}{}{\text{B}}}{\LocalIISeq{1}{}{\text{B}}} \right)^{-1}
        > \frac{c_{10} q_2^{\ast}}{c_{01} (1 - q_2^{\ast})} \frac{\LocalISeq{1}{}{\text{B}}}{1 - \LocalIISeq{1}{}{\text{B}}}.
    \end{align}
    The terms in the left-hand and the right-hand sides are the same as the belief update formulae~\eqref{eq:BeliefUpdate2-0} and~\eqref{eq:BeliefUpdate2-2}, so this simplifies to
    \begin{equation}
        \left( \frac{c_{10} \Belief_2^{_0}}{c_{01} (1 - \Belief_2^{_0})} \right) ^{-1} > \frac{c_{10} \Belief_2^{_1}}{c_{01} (1 - \Belief_2^{_1})}.
        \label{eq:SlopeComparison1}
    \end{equation}

    Let us discuss the meaning of the inequality~\eqref{eq:SlopeComparison1}.  In Fig.~\ref{fig:ROCCurve}, the convex curve depicts a flipped version of the receiver operating characteristic (ROC).  When the prior belief is $\Belief$, the error probabilities $(\LocalISeq{2}{}{}, \LocalIISeq{2}{}{})$ are determined as the point of tangency, where the curve meets a line with slope $- c_{10} \Belief / c_{01} (1 - \Belief)$.  Two solid lines in Fig.~\ref{fig:ROCCurve} depict the lines for Blake's updated beliefs after observing $\Hhat_1 = 0$ and $\Hhat_1 = 1$, respectively denoted by $\Belief_2^{_0}$ and $\Belief_2^{_1}$.

    \begin{figure}
        \centering
        \includegraphics[width=3.2in]{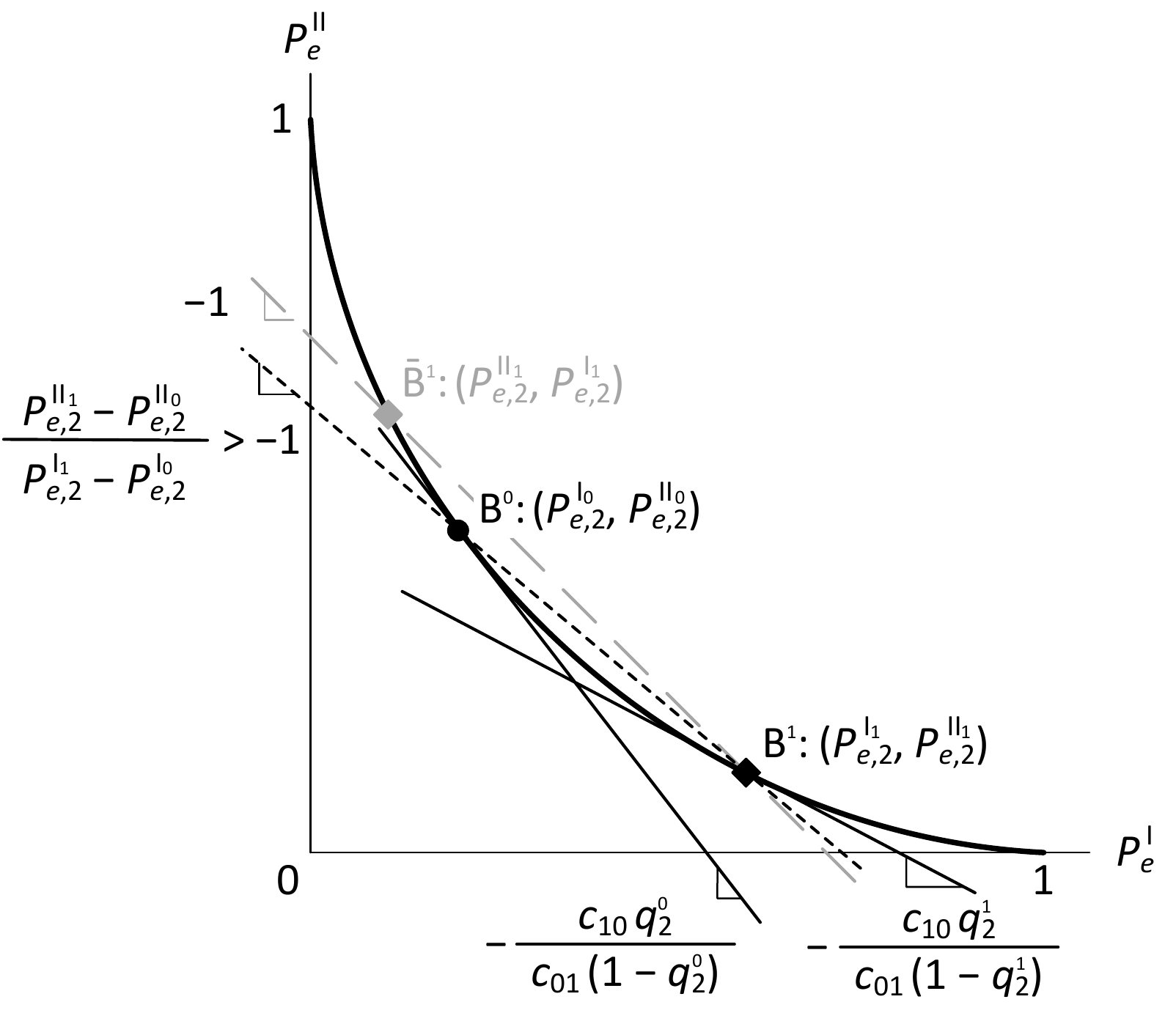}
        \caption{The point $\text{B}^{0}$ $(\LocalISeq{2}{0}{}, \LocalIISeq{2}{0}{})$ always exists between the points $\text{\={B}}^{1}$ $(\LocalIISeq{2}{1}{}, \LocalISeq{2}{1}{})$ and $\text{B}^{1}$ $(\LocalISeq{2}{1}{}, \LocalIISeq{2}{1}{})$.}
        \label{fig:ROCCurve}
    \end{figure}

    The inequality~\eqref{eq:SlopeComparison1} restricts the range of error probabilities in which $(\LocalISeq{2}{0}{}, \LocalIISeq{2}{0}{})$ can exist on the basis of $(\LocalISeq{2}{1}{}, \LocalIISeq{2}{1}{})$; the point $\text{B}^{_0}$ $(\LocalISeq{2}{0}{}, \LocalIISeq{2}{0}{})$, a black dot, always exists on the right side of the point $\bar{\text{B}}^{_1}$ $(\LocalIISeq{2}{1}{}, \LocalISeq{2}{1}{})$, a gray diamond. 
    Furthermore, the point $\text{B}^{_0}$ cannot exist on the right side of the point $\text{B}^{_1}$ $(\LocalISeq{2}{1}{}, \LocalIISeq{2}{1}{})$, a black diamond, because obviously $\Belief_2^{_0} > \Belief_2^{_1}$.  Therefore, the point $\text{B}^{_0}$ always exists on the curve between the points $\bar{\text{B}}^{_1}$ and $\text{B}^{_1}$.

    Now we draw a black dotted line that connects the points $\text{B}^{_0}$ and $\text{B}^{_1}$ and a gray dashed line that connects the points $\bar{\text{B}}^{_1}$ and $\text{B}^{_1}$.  From the restriction for the point $\text{B}^{_0}$, the slope of the former is always greater than that of the latter:
    \begin{equation}
        \frac{\LocalIISeq{2}{1}{} - \LocalIISeq{2}{0}{}}{\LocalISeq{2}{1}{} - \LocalISeq{2}{0}{}} > -1.
        \label{eq:SlopeComparison2}
    \end{equation}

    We have obtained the optimality condition \eqref{eq:Derivative1-1} for Alexis's prior belief in Section~\ref{sec:OptimalInitialBelief}.  We can rewrite it as
    \begin{equation}
        \Belief_1^{\ast} = \frac{p_0}{p_0 + (1 - p_0) \frac{\LocalIISeq{2}{0}{} - \LocalIISeq{2}{1}{}}{\LocalISeq{2}{1}{} - \LocalISeq{2}{0}{}}}.
    \end{equation}
    Finally, we can conclude that $\Belief_1^{\ast} > p_0$ because of~\eqref{eq:SlopeComparison2}.

    In addition, Alexis's optimal belief $\Belief_1^{\ast}$ is upper-bounded by $\Frac{c_{01}}{(c_{10} + c_{01})}$ because $\Belief_1^{\ast}$ is strictly decreasing in $p_0$ and $\Belief_1^{\ast} = \Frac{c_{01}}{(c_{10} + c_{01})}$ when $p_0 = \Frac{c_{01}}{(c_{10} + c_{01})}$ by Theorem~\ref{thm:TrueIsOptimal}.
    Combining these two bounds, we have the inequality
    \begin{equation}
        p_0 < \Belief_1^{\ast} < \frac{c_{01}}{c_{10} + c_{01}},
    \end{equation}
    as desired.

    The statement that $\Frac{c_{01}}{(c_{10} + c_{01})} < \Belief_1^{\ast} < p_0$ if $p_0 > \Frac{c_{01}}{(c_{10} + c_{01})}$ can be proven similarly.
\end{IEEEproof}

\section*{Acknowledgment}
Discussions with V. Krishnamurthy, J. Z. Sun, V. Montazerhodjat, and L. R. Varshney are greatly appreciated.

\bibliographystyle{IEEEtran}
\bibliography{rhim_lib}
\end{document}